%% file: preprint_Arxiv_v2.tex
\documentclass{article}

\usepackage{PRIMEarxiv}

\usepackage[utf8]{inputenc} 
\usepackage[T1]{fontenc}    
\usepackage{url}            
\usepackage{booktabs}       
\usepackage{amsfonts}       
\usepackage{nicefrac}       
\usepackage{microtype}      
\usepackage{lipsum}
\usepackage{fancyhdr}       
\usepackage{graphicx}       

\input{utils/packages}
\input{utils/defined_colors_and_commands}

\usepackage{hyperref}
\definecolor{pastelblue}{RGB}{90,130,190}
\hypersetup{
    colorlinks=true,
    linkcolor=pastelblue,
    citecolor=pastelblue,
    urlcolor=pastelblue,
    filecolor=pastelblue,
    pdfborder={0 0 0}
}

\graphicspath{{media/}}     

\title{Optimization-Embedded Active Multi-Fidelity Surrogate Learning for Multi-Condition Airfoil Shape Optimization}

\author{
  Isaac Robledo \\
  Aerial Platforms Department \\ Subdirectorate General for Aeronautical Systems\\ 
  Spanish National Institute for Aerospace Technology (INTA)\\ 
  \vspace{-0.2cm}\\
  Department of Aerospace Engineering \\
  Universidad Carlos III de Madrid \\
  Madrid, Spain \\
  \texttt{isaac.robledo@alumnos.uc3m.es} \\
   \And
  Alberto Vilari\~no \\
  Department of Aerospace Engineering \\
  Universidad Carlos III de Madrid \\
  Madrid, Spain \\
  \texttt{alberto.vilarino@alumnos.uc3m.es} \\
  \And
  Arnau Mir\'o \\
  TUAREG Research Group\\
  Universitat Politècnica de Catalunya (UPC)\\
  \vspace{-0.2cm}\\
  CASE Large-Scale Computational Fluid Dynamics\\
  Barcelona Supercomputing Centre (BSC)\\
  Barcelona, Spain \\
  \texttt{arnau.miro@upc.edu} \\
  \And
  Oriol Lehmkuhl \\
  CASE Large-Scale Computational Fluid Dynamics\\
  Barcelona Supercomputing Centre (BSC)\\
  Barcelona, Spain \\
  \texttt{oriol.lehmkuhl@bsc.es} \\
  \And
  Carlos Sanmiguel Vila\\
  Aerial Platforms Department \\ Subdirectorate General for Aeronautical Systems\\ 
  Spanish National Institute for Aerospace Technology (INTA)\\
  \vspace{-0.2cm}\\
  Department of Aerospace Engineering\\
  Universidad Carlos III de Madrid \\
  Madrid, Spain \\
  \texttt{csanvil@inta.es} \\
  \And
  Rodrigo Castellanos \\
  Department of Aerospace Engineering \\
  Universidad Carlos III de Madrid \\
  Madrid, Spain \\
  \texttt{rcastell@ing.uc3m.es} \\
}

\begin{document}
\maketitle
\vspace{-1cm}
\begin{abstract}
Active multi-fidelity surrogate modeling is developed for multi-condition airfoil shape optimization to reduce high-fidelity CFD cost while retaining RANS-consistent aerodynamic metrics. The framework couples a low-fidelity-informed Gaussian process regression transfer model with uncertainty-triggered sampling and a synchronized elitism rule embedded in a hybrid genetic algorithm. Low-fidelity XFOIL evaluations provide inexpensive features, while sparse RANS simulations are adaptively allocated when predictive uncertainty exceeds a threshold;  elite candidates are mandatorily validated at high fidelity, and the population is re-evaluated to prevent evolutionary selection based on outdated fitness values produced by earlier surrogate states. The method is demonstrated for a two-point problem at $Re=6\times10^6$ with cruise at $\alpha=2^\circ$ (maximize $E=L/D$) and take-off at $\alpha=10^\circ$ (maximize $C_L$) using a 12-parameter CST representation. Independent multi-fidelity surrogates per flight condition enable decoupled refinement. The optimized design improves cruise efficiency by 41.05\% and take-off lift by 20.75\% relative to the best first-generation individual. Over the full campaign, RANS evaluations were required for only 14.78\% and 9.5\% of the condition-specific candidate evaluations at cruise and take-off, respectively. These percentages quantify the reduction in high-fidelity usage relative to the fixed automated RANS workflow adopted as the high-fidelity reference in this study.
\end{abstract}


\twocolumn 

\section{Introduction}
Advances in computational aerodynamics have made automated shape optimization a viable component of modern design workflows. Compared with traditional cycles dominated by wind-tunnel testing and empirical iteration, CFD-driven optimization enables systematic exploration with fewer simplifying assumptions. Its practical deployment, however, is constrained by the high dimensionality typical of aerodynamic design spaces and by the cost of repeated high-fidelity evaluations. These constraints motivate surrogate-based global optimization frameworks that manage fidelity adaptively, combining low-cost models with targeted high-fidelity corrections.

Historically, shape optimization has relied on gradient-based methods coupled with adjoint equations \citep{jameson1988aerodynamic, reuther1996aerodynamic}. Although computationally efficient in terms of design-variable scaling, these approaches typically require a tight integration between the flow solver and the optimizer, which often leads to problem-specific implementations and limits portability in industrial settings \citep{slotnick2014cfd}. 
To overcome these limitations, the field has increasingly adopted surrogate modeling. Trained on an initial set of simulations, surrogate models provide inexpensive predictions across the design space, which facilitates the use of global, derivative-free optimizers \citep{yondo2018review}. Among the various surrogate options, Kriging remains one of the most widely used surrogate choices in aerodynamic design \citep{han2012hierarchical, nagawkar2020applications}. However, its standard formulations can become computationally expensive to train and deploy for high-dimensional inputs or large datasets \citep{laurenceau2008building}. To mitigate these costs, Kriging is typically coupled with gradient information \citep{han2017weighted, bouhlel2019gradient} or employs partial least squares to reduce training costs \citep{bouhlel2016improving}.

Building on these traditional frameworks, recent advancements in surrogate-based optimization have integrated machine learning models as high-capacity physics evaluators. This shift aims to further reduce reliance on repeated CFD simulations, offering substantial gains in evaluation speed, albeit often at the expense of fidelity relative to high-fidelity solvers. Representative examples include fully connected neural networks \citep{wu2023airfoil} or convolutional neural networks \citep{Esfahanian2023shapeoptcnn} for aerodynamic coefficient prediction, Physics-Informed Neural Networks (PINNs) for physics-constrained inference \citep{raissi2019physics}, and DeepONets \citep{shukla2023deep}, which enable mesh-independent operator learning \citep{lu2021learning}. More advanced methodologies lean towards merging these techniques; for example, \citet{Pereira2025supersonicairfoilshapeopt} couples Generative Adversarial Network (GAN) based parameterization with CNN coefficient prediction for supersonic airfoil shape optimization, while \citet{liu2025cnn} uses a deep reinforcement learning (DRL) procedure to integrate CNN-based parametric extraction with PINN aerodynamic evaluation. 

Parallel to these advancements in predictive modeling, the problem of geometric parametrization remains a critical bottleneck, as the choice of representation directly dictates the design-space dimensionality and the feasibility of generated shapes. In airfoil optimization, classical methods have long relied on B-spline and B\'ezier representations \citep{rajnarayan2018universal}, PARSEC \citep{della2014airfoil}, Hicks-Henne bump functions \citep{he2019improved}, and shape-class transformation (CST) \citep{kulfan_universal_2007, vu2015aerodynamic, zhang2015supercritical}. 
More recently, data-driven methods have introduced dimensionality reduction and generative modeling into airfoil parametrization. Approaches range from projection-based solutions using Proper Orthogonal Decomposition (POD) \citep{toal2010geometric, wu2019benchmark, li2019data} to nonlinear latent-variable model techniques such as Variational Autoencoders (VAE) \citep{yonekura2021data} and GAN \citep{achour2020development}. Other strategies focus on learning parametrizations directly, for example, through GAN-based shape generation \citep{chen2020airfoil}, or on extracting compact parametric descriptions from existing data \citep{sekar2019fast}.

Both surrogate modeling and shape parametrization aim to mitigate the inherent complexity, non-convexity, and high dimensionality of aerodynamic shape optimization, either by reducing the effective design space through parametrization or by accelerating evaluations through surrogate modeling. Multi-fidelity techniques address this challenge from a complementary perspective by combining information sources of differing cost and reliability. In this paradigm, a large number of inexpensive low-fidelity evaluations are exploited to explore the design space, while a limited set of high-fidelity simulations is employed to correct their deficiencies, enabling improved performance estimates at reduced computational cost \citep{Giselle_Fern_ndez_Godino_2023_mfreview,Schouler2025costcomparison}. 

To implement this multi-fidelity paradigm, research has traditionally branched into two distinct architectural strategies. The first involves hierarchical approaches, which selectively invoke different fidelity levels depending on the design stage or region of interest; however, these can incur increasing algorithmic complexity and reduced interpretability as the number of information sources grows \citep{Giselle_Fern_ndez_Godino_2023_mfreview}. In contrast, surrogate-based multi-fidelity models seek to fuse available information into a unified predictive framework. Because this approach offers a balance of accuracy and efficiency, it serves as the basis for the methodology developed in this study. Early formulations relied on additive or multiplicative corrections to low-fidelity predictions, $\hat{y}_{HF}(x) = \rho(x)\,y_{LF}(x) + \delta(x)$, where $\rho$ and $\delta$ are fitted correction terms. Representative examples in aeronautics include the additive model of \citet{Alexandrov2001mfadditive} and the multiplicative formulation of \citet{Balabanov1999mfmultiplicative}, later extended through more flexible weighting and regression strategies \citep{Zheng2013mfadvanceweighting,fischer2017mfweighting}.  

With the adoption of modern machine-learning techniques, multi-fidelity surrogate modeling has received increasing attention in aerodynamics, where the non-linearity of the governing equations and the diversity of available fidelity sources (from panel methods to RANS, LES, DNS, and experimental data) challenge classical correction strategies \citep{schouler2025costbayesian, Schouler2025costcomparison}. Among these approaches, co-Kriging is a widely used framework, as it can exploit statistical correlations between fidelity levels to construct improved predictors \citep{Qian2008cokrigin, FORRESTER2009cokrigin, Gratiet_2014_mf, BREVAULT2020_mf}. Deep learning has also been used to construct multi-fidelity models, including multi-fidelity neural networks \citep{MENG202010mfnn}, multi-fidelity deep Bayesian neural networks \citep{MENG2021mfbayesiannn}, multi-fidelity deep Gaussian processes \citep{cutajar2019deepgaussianprocessesmultifidelity}, multi-fidelity deep neural operators \citep{HOWARD2023neuralmf}, and multi-fidelity reinforcement learning \citep{BHOLA2023mfrl}. For the interested reader, a broader survey on multi-fidelity surrogate models for optimization is provided by \citet{li2024multifidelitymethodsoptimizationsurvey}.

However, despite their success, many existing multi-fidelity surrogates are constructed offline and subsequently assumed to remain valid throughout the optimization process. This assumption becomes fragile in large, weakly constrained design spaces, where optimizers push the search toward unexplored regions and potentially distinct physical regimes. Under these conditions, the correlation between low- and high-fidelity models may locally deteriorate, and static correction models can rapidly lose reliability. These limitations motivate multi-fidelity strategies that not only fuse heterogeneous information sources but also adapt dynamically as the optimization progresses.

Active learning strategies address the limitations of static surrogate models by iteratively adapting them through the targeted acquisition of new training data. In this context, Bayesian optimization is a prominent example, where uncertainty-aware acquisition functions guide sample selection to improve model accuracy while supporting optimization. Motivated by this paradigm, extensive research has focused on integrating active learning into multi-fidelity aerodynamic optimization frameworks. Representative examples include partitioning candidate designs between low- and high-fidelity evaluators within an optimization loop \citep{ZHANG2021mfdnn}, uniform infilling strategies embedded in multi-fidelity shape optimization \citep{MOHAMMADZADEH2018mf}, and expected-improvement-driven enrichment in co-Kriging-based airfoil optimization \citep{AYE2023multiobjEIcokriging}. More generally, adaptive infill strategies have shown strong performance by selecting additional samples aimed at improving surrogate reliability in the region of interest \citep{ZHANG2021mfadaptiveinfill, CHARAYRON2023mfadaptiveinfill, WU2024mfadaptiveinfill, MOUROUSIAS2024mfadaptiveinfill}.

However, in many of these formulations, active learning is primarily posed as a surrogate-enrichment problem: additional high-fidelity data is selected to improve the model, while the optimization process that generates and ranks candidate designs remains only loosely coupled to that enrichment decision. In practice, high-fidelity evaluations are typically allocated through predefined schedules, fixed per-iteration budgets, or acquisition rules applied globally. Such strategies are effective for improving surrogate quality, but they do not necessarily decide whether a specific candidate currently being promoted by the optimizer is reliable enough to be selected without high-fidelity confirmation. 
As a result, these strategies may not adapt fidelity selection to the local reliability of the surrogate along the optimizer trajectory, and they only indirectly account for the evolving structure of the optimization landscape. In large, multi-modal design spaces, where the search progressively concentrates around high-performing regions and may encounter distinct physical regimes, such enrichment policies can lead to inefficient allocation of computational resources and delayed correction of surrogate inaccuracies. The key issue is therefore not only where new high-fidelity information should be added, but also when fidelity should be escalated during candidate evaluation and how the optimization population should be kept consistent after the surrogate has been updated.

This distinction is relevant with respect to classical sequential Kriging, adaptive Kriging, and Bayesian-optimization-type infill methods. In those approaches, the surrogate update is commonly driven by a separate acquisition or learning-function problem. Here, by contrast, the high-fidelity decision is embedded directly in the evaluation of candidates generated by the evolutionary optimizer: fidelity escalation is performed condition-wise at the candidate level, and subsequent surrogate updates are coupled with high-fidelity elite validation and synchronized population re-evaluation.

The objective of this work is therefore not to introduce a new acquisition function or to provide an exhaustive benchmark against all possible active-learning policies. Instead, the focus is on the optimizer-embedded use of fidelity escalation and synchronization mechanisms needed when a multi-fidelity surrogate is updated during an evolutionary search. The validation is consequently designed to isolate whether online refinement, elite high-fidelity validation, and population re-evaluation prevent surrogate-driven selection errors relative to a static LF-informed surrogate baseline.

To address this gap, we propose an optimization-embedded active multi-fidelity framework for multi-condition airfoil shape optimization that couples surrogate modeling, uncertainty quantification, and evolutionary search within a unified process. The core mechanism is an uncertainty-triggered high-fidelity evaluation strategy: rather than prescribing high-fidelity calls a priori, expensive simulations are allocated dynamically based on the local reliability of the evolving multi-fidelity surrogate. High-fidelity evaluations are therefore invoked only when needed, enabling iterative correction of the surrogate as the optimizer explores new regions of the design space.  
Furthermore, the framework introduces a condition-wise decoupling of the multi-fidelity models, allowing the optimization to simultaneously account for multiple flight conditions, each with its own uncertainty structure and refinement requirements. This feature is particularly relevant in multi-point aerodynamic design, where the correlation between low- and high-fidelity models may differ substantially across operating regimes.  

Multi-fidelity fusion is here formulated as a data-driven transfer mapping rather than a hierarchical autoregressive model. Specifically, the high-fidelity response is learned as a nonparametric function of the design variables and the associated low-fidelity outputs, which avoids assuming a globally valid linear correlation between fidelity levels. Active learning is then used to introduce high-fidelity corrections in regions where the low-to-high relationship degrades along the optimization trajectory.

The proposed active multi-fidelity strategy is embedded within a hybrid global evolutionary optimizer, HyGO \citep{robledo2025hygo}, which provides synchronization between the evolving population and the adaptive fidelity-selection process. This integration supports robust global search over large, weakly constrained design spaces, while ensuring consistency between the surrogate models, the objective evaluations, and the evolutionary dynamics. The implementation in HyGO removes the need for fixed per-iteration high-fidelity budgets by making fidelity escalation a candidate-level decision driven by surrogate uncertainty, and it includes an elite-consistency mechanism to avoid selection drift after surrogate updates.

The proposed approach is demonstrated on a two-condition airfoil shape optimization problem using a 12-parameter CST representation, combining XFOIL and RANS simulations. The results provide quantitative evidence of surrogate convergence during optimization, reductions in high-fidelity cost through adaptive refinement, and physically interpretable aerodynamic trends, illustrating the behavior of the proposed strategy on a multi-condition airfoil design problem.

The main contributions of this work are threefold: 
(i) an optimization-embedded active multi-fidelity refinement strategy that couples candidate-level uncertainty-triggered high-fidelity escalation with mandatory high-fidelity validation of elite individuals and synchronized population re-evaluation within the evolutionary search; 
(ii) a condition-wise low-fidelity-informed Gaussian-process transfer formulation, in which independent surrogates map geometric variables and low-fidelity aerodynamic outputs to high-fidelity-consistent predictions for each operating point; and
(iii) a two-condition CST-based airfoil optimization study combining XFOIL low-fidelity evaluations and RANS high-fidelity simulations, including surrogate-error reduction, high-fidelity budget analysis, and additional ablation/validation analyses designed to isolate the contribution of the active refinement mechanism.

The remainder of this paper is organized as follows. \hyperref[s:methodology]{Section~\ref*{s:methodology}} presents the proposed methodology, including the airfoil parametrization, the numerical setup of the low- and high-fidelity solvers, the multi-fidelity surrogate modeling strategy, and the optimization-embedded active learning framework. \hyperref[s:Results]{Section~\ref*{s:Results}} discusses the results of the multi-condition airfoil optimization campaign, including convergence behavior, computational cost analysis, surrogate validation, and physical interpretation of the optimized geometries. Finally, \hyperref[s:Conclusions]{Section~\ref*{s:Conclusions}} summarizes the main conclusions and outlines perspectives for future developments.

\section{Methodology}\label{s:methodology}

This section details the methodology proposed for the multi-fidelity airfoil optimization framework. The discussion is organized into four main components: the geometric parametrization used to define the airfoil shape, the numerical setup of the low- and high-fidelity CFD solvers, the definition of the multi-fidelity surrogate environment, and the genetic algorithm-based optimization strategy.

\subsection{Airfoil Parametrization}\label{ss:parametrization}

\begin{table}
    \caption{\label{tab:CST_domain} Parametric domain for the 12 parameter CST representation.}
    \centering
    \begin{tabular}{c|cc}
        \hline
        \textbf{Parameter} & \textbf{Upper Limit} & \textbf{Lower Limit} \\ 
        \hline\hline
        $\theta_1$    & 0.60377 &  0.03834 \\
        $\theta_2$    & 0.70389 & -0.35002 \\
        $\theta_3$    & 0.79453 & -0.53920 \\
        $\theta_4$    & 0.81549 & -0.42392 \\
        $\theta_5$    & 0.89804 & -1 \\
        $\theta_6$    & 1       & -1 \\
        $\theta_7$    & 0.29979 & -0.38389 \\
        $\theta_8$    & 0.53284 & -0.55331 \\
        $\theta_9$    & 0.36506 & -0.77029 \\
        $\theta_{10}$ & 0.60049 & -0.62099 \\
        $\theta_{11}$ & 0.38495 & -0.79700 \\
        $\theta_{12}$ & 0.86272 & -1 \\ 
        \hline\hline
    \end{tabular}
\end{table}

Airfoil shape parametrization is a pivotal aspect of any shape optimization framework, as it defines the parametrization and dimensionality of the design space. This study employs the CST methodology \citep{kulfan_universal_2007} for its flexibility, robustness, and ability to represent a broad range of shapes with a concise set of parameters. The airfoil geometry is defined by a class function (defining the general airfoil shape) and a shape function (typically Bernstein polynomials). The coefficients of these polynomials serve as the design variables. For this study, a 12-parameter representation was selected (six coefficients for the upper surface, six for the lower), offering a trade-off between geometric flexibility and low dimensionality for the optimization algorithm.

\begin{table}[]
\centering
\caption{\label{tab:CST_constraints} Geometrical constraints for CST-generated airfoils with unitary chord}
\resizebox{8.cm}{!}{
    \begin{tabular}{@{}clc@{}}
    \toprule
    Target & Constraint & Valid Range \\ \midrule
    \multirow{3}{*}{\begin{tabular}[c]{@{}c@{}}Global \\ geometric \\ bounds\end{tabular}} & Maximum thickness & 5\%-50\% \\
     & Maximum thickness $x$-position & 20\%-35\% \\
     & Contiguous panels angle in upper surface & $<10^\circ$ \\ \midrule
    \multirow{5}{*}{\begin{tabular}[c]{@{}c@{}}TE\\ geometry\end{tabular}} & Minimum thickness in TE vicinity & 0.2\% \\
     & Minimum $x$ for collapsed thickness & 96\% \\
     & Minimum thickness region & 88\%-96\% \\
     & Minimum TE angle & $3^\circ$ \\
     & Monotonicity in TE region & True \\ \midrule
    \begin{tabular}[c]{@{}c@{}}Undesired \\ shapes\end{tabular} & Minimum airfoil area & $0.02\cdot t_{max}$ \\ \bottomrule
    \end{tabular}}
\end{table}

Defining the parametric domain (i.e., the bounds of the design variables) is critical. To ensure the domain encompasses realistic aerodynamic shapes, a broad reference database spanning multiple airfoil families was assembled, including the NACA 4- and 5-digit series, and representative profiles from NACA 6-series, RAE, DAE, RAG, Boeing, Whitcomb, NLF, ONERA, and the Eppler E398 and SC families. All airfoils were then fitted using the CST formulation. This procedure reproduces established aerodynamic shapes while allowing for intermediate geometries beyond the original family definitions. To balance exploration and feasibility, the CST bounds inferred from this dataset were expanded by 15\% on both sides. 
The resulting parametric domain is detailed in \autoref{tab:CST_domain}, where $\theta_1$-$\theta_6$ correspond to the upper surface and $\theta_7$-$\theta_{12}$ to the lower surface, and $\boldsymbol{\Theta}=[\theta_1,\dots,\theta_{12}]$ is the parametric vector characterizing a specific airfoil shape. Despite this bounded domain, specific coefficient combinations may still yield degenerate geometries (e.g., self-intersecting surfaces or zero-thickness regions). To mitigate this, a set of geometric validity constraints is enforced, so that invalid airfoils are treated via a death-penalty rule as detailed in \autoref{ss:Cost_function}.
\begin{figure}[htb]
    \centering
    \includegraphics[width=8.cm]{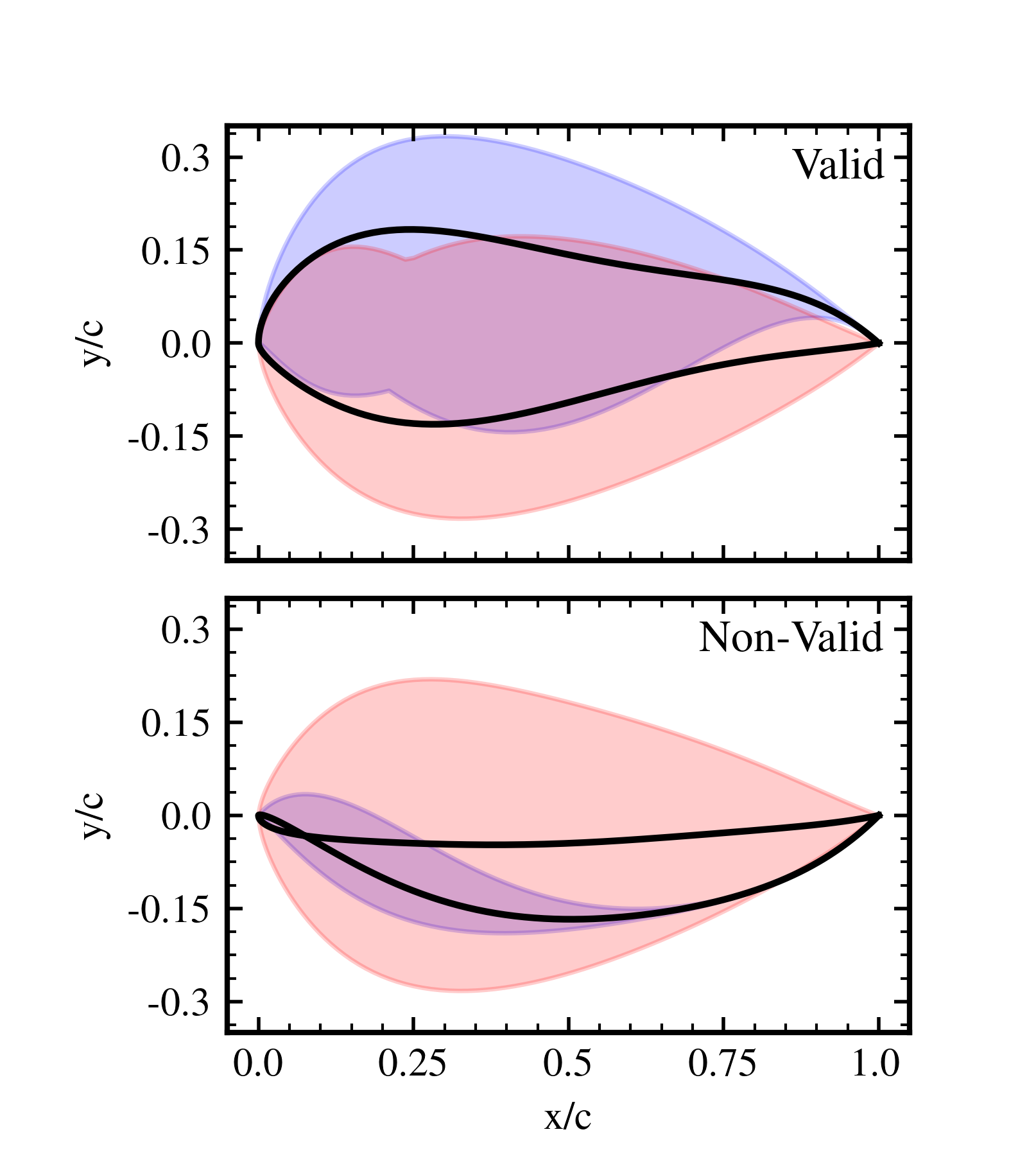}
    \caption{CST airfoil design envelope from three-level full-factorial sampling of \autoref{tab:CST_domain}: valid geometries (top), rejected geometries (bottom), mean surface, and upper/lower bounds.}
    \label{fig:01_desig_space_CST}
\end{figure}

The constraints, summarized in \autoref{tab:CST_constraints}, are defined for normalized chord coordinates ($x/c, y/c$). Thickness constraints ensure structural feasibility, while curvature constraints (based on the angle between contiguous panels) enforce smoothness on the upper surface to prevent non-physical oscillations. The lower surface is granted higher geometric freedom to allow for the complex curvature commonly associated with supercritical airfoils. Finally, the trailing edge (TE) region is strictly constrained to prevent premature thickness collapse; specifically, the airfoil thickness in the last 12\% is enforced to remain above 0.2\% until the final 4\% of the chord, ensuring a realistic TE wedge angle.

To demonstrate the effectiveness of these constraints in filtering invalid geometries, a discrete sampling of the parametric space was performed using three equidistant values per design variable. \autoref{fig:01_desig_space_CST} visualizes the resulting design envelope, distinguishing between valid and invalid geometries with color-coded boundaries. The results indicate that the 12-parameter CST formulation yields a broad design space capable of representing diverse aerodynamic shapes, while the imposed constraints effectively segregate degenerate or non-physical curves from the feasible population.

\subsection{Numerical Setup}
The optimization framework is designed to evaluate aerodynamic performance across varying levels of fidelity. The design problem targets airfoils operating in the incompressible regime at a Reynolds number of $Re=6\times10^6$. Two distinct flight conditions are selected to drive the optimization: a cruise condition at an angle of attack $\alpha=2^\circ$, representing the primary operational phase, and a take-off condition at $\alpha=10^\circ$. The latter serves as a critical high-lift case, essential for aircraft configurations where complex high-lift devices may be constrained by weight or mechanical complexity.

The low-fidelity aerodynamic evaluations are performed using XFOIL \citep{drela1989xfoil}. XFOIL is a standard tool for airfoil design that couples a high-order inviscid panel method with a fully interactive integral boundary-layer formulation. This viscous-inviscid interaction approach enables rapid prediction of lift and drag with reasonable accuracy in attached-flow conditions \citep{morgado2016xfoil,zhang2019xfoil,carreno2024experimental}. As with other integral boundary-layer approaches, convergence can degrade near stall or for geometries that induce strong separation; such cases are treated as invalid low-fidelity evaluations within the optimization workflow. 

\begin{figure}[htb]
    \centering
    \includegraphics[width=8.cm]{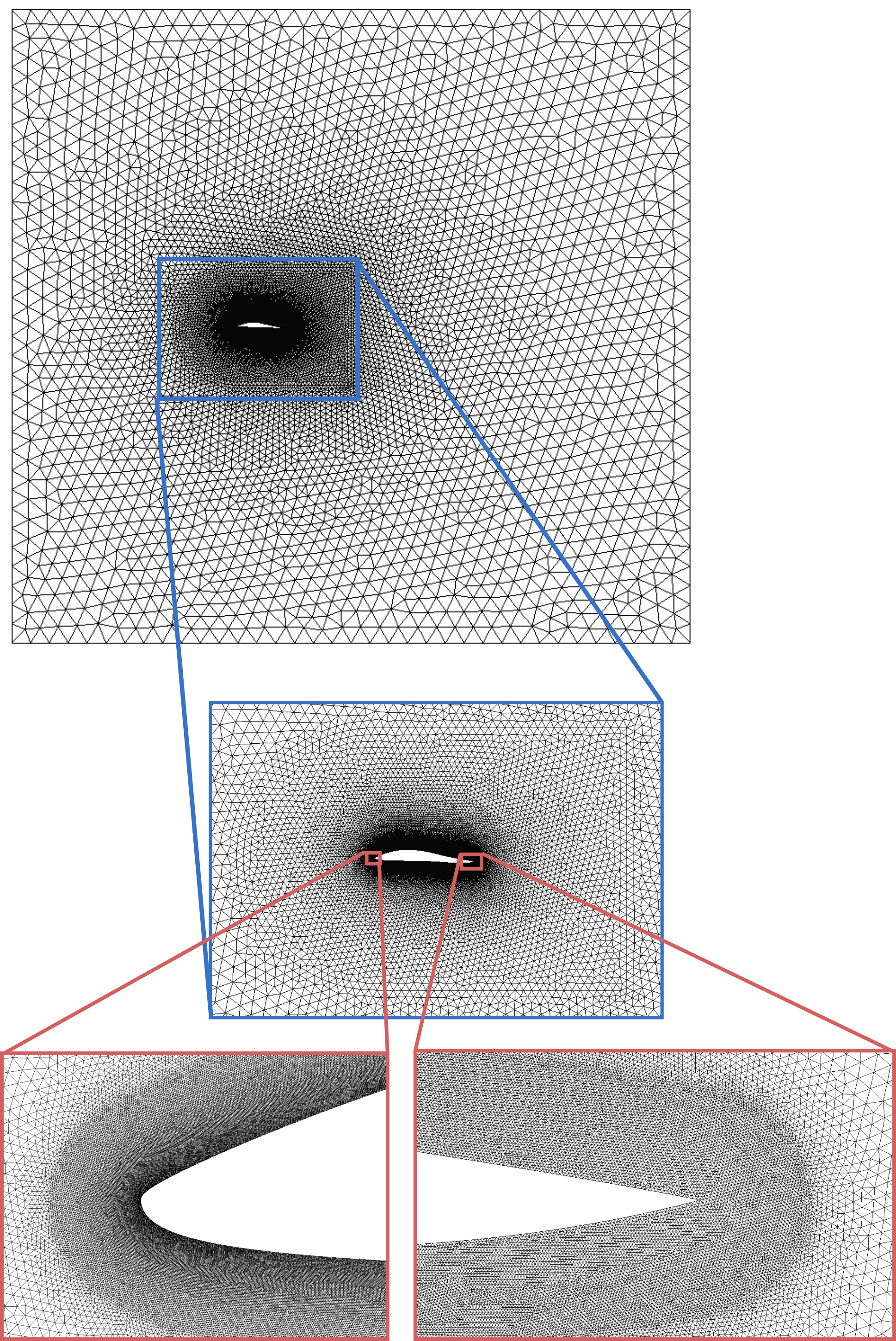}
    \caption{Computational grid for a representative airfoil geometry. The visualization highlights the unstructured domain discretization, the local refinement zones, and the inflation layers resolving the boundary layer at the leading and trailing edges.}
    \label{fig:02_mesh}
\end{figure}

High-fidelity aerodynamic simulations are performed using the open-source CFD toolbox OpenFOAM \citep{JASAK2009openfoam}. The incompressible transient finite-volume solver \texttt{pimpleFoam}, which utilizes the PIMPLE algorithm (a combination of PISO and SIMPLE), was employed using a first-order pseudo-transient time scheme designed for steady cases to solve the Reynolds-Averaged Navier-Stokes (RANS) equations. Turbulence was modeled using the $k-\omega$ Shear Stress Transport (SST) model \citep{menter1994two}, with boundary conditions as detailed in \autoref{tab:RANS_BC}. The computational domain spans $[-5c, 10c]$ in the streamwise direction and $[-7c, 7c]$ in the vertical direction, normalized by the airfoil chord $c=1$, and with origin at the airfoil Leading Edge (LE). The domain is discretized using an unstructured triangular-element grid. To resolve flow gradients effectively, local mesh refinement is applied in two zones: a circular region of radius $R = 1.5c$ centered at the LE, and a rectangular wake region extending across $[0, 2.3c] \times [-1.5c, 1.5c]$. Each RANS case was advanced in pseudo-time until the normalized residuals of all solved variables decreased below $10^{-6}$, with a maximum limit of 50,000 iterations. Cases reaching this limit were accepted only when the final aerodynamic-coefficient histories satisfied the prescribed stationarity and admissibility checks described in \autoref{ss:Optimization_framework}.
To capture boundary layer effects while managing computational cost, an inflation layer with a total height of $0.0352c$ was generated around the airfoil surface. The first cell height was sized to yield an average non-dimensional wall distance of $y^+ \approx 40$, with a maximum of $y^+ \approx 80$. This resolution is consistent with the wall-function treatment used in the RANS setup. \autoref{fig:02_mesh} illustrates the mesh topology and refinement zones described above.

As an a posteriori verification of the automated meshing pipeline, the final optimized geometry (presented in \autoref{s:Results}) was also inspected at both operating conditions. The resulting near-wall resolution remained compatible with the wall-function treatment: at $\alpha=2^\circ$, the minimum, average, and maximum values were $y^+=1.64$, $34.2$, and $72$, respectively, while at $\alpha=10^\circ$ they were $y^+=0.95$, $33$, and $76$. The corresponding mesh-quality diagnostics also remained within acceptable OpenFOAM limits, with a maximum aspect ratio of $2.48$, a maximum non-orthogonality of $31.1^\circ$ (average $3.82^\circ$), and a maximum skewness of $0.529$. Cell openness, face areas, cell volumes, face pyramids, and coupled-point consistency checks were all reported as valid. These diagnostics indicate that the same automated unstructured near-wall refinement strategy remained robust when applied to the most relevant optimized geometry, without requiring manual mesh repair or topology-specific intervention.

\begin{figure}[htb]
    \centering
        \includegraphics[width=8.cm]{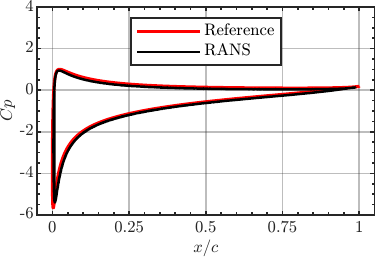}
        \caption{Surface pressure of the finest mesh versus the reference surface pressure, obtained from \citet{jespersen2016overflowReference}.}
        \label{fig:00_convergence}
\end{figure}

To ensure numerical robustness, a mesh convergence study was performed at the most critical condition, corresponding to take-off at $\alpha=10^\circ$. A total of five mesh refinements were evaluated. The results, presented in \autoref{tab:mesh_convergence}, demonstrate a convergence towards the reference data \citep{jespersen2016overflowReference}, achieving errors of just 1.26\% for $C_D$ and 0.24\% for $C_L$ on the finest mesh (the reference values correspond to the average across SST implementations reported in \citet{jespersen2016overflowReference}). Furthermore, \autoref{fig:00_convergence} illustrates the excellent agreement of the surface pressure distribution between the reference and the computed results for the finest mesh. Although this mesh configuration is denser than those typically used in industrial shape optimization, the extended design space defined by the CST parametrization may lead to automatically generated geometries with highly distorted or irregular shapes. Therefore, the finest mesh level reported in Table~\ref{tab:mesh_convergence} is retained as a fixed high-fidelity reference resolution for all RANS evaluations to ensure robust and reliable simulations. The purpose of this choice was not to identify the minimum-cost RANS discretization for a single airfoil, but to define a robust and repeatable automated CFD pipeline that could be applied without manual mesh intervention across the CST design space. This is important for the present multi-fidelity study because the high-fidelity source must remain consistent when generating training data, validating elite candidates, and estimating the computational savings associated with avoided RANS calls. Consequently, the reported cost reduction should be interpreted relative to this fixed high-fidelity reference workflow. A different RANS discretization or solver setup would change the absolute computational cost, but not the algorithmic definition of the proposed fidelity-escalation and synchronization strategy.

\begin{table}
    \caption{\label{tab:RANS_BC} Boundary conditions for high-fidelity RANS simulations. Specific initialization or fixed values are enclosed in brackets $[\cdot]$. $\omega$ and $k$ values follow the recommendations by \citet{1992bookMenter}}
    \centering
    \scriptsize 
    \begin{tabularx}{8.2cm}{@{} >{\bfseries}l >{\raggedright\arraybackslash}X >{\raggedright\arraybackslash}X >{\raggedright\arraybackslash}X >{\raggedright\arraybackslash}X @{}}
    \hline\hline
    Field & Inlet & Outlet & Top/Bot. & Airfoil \\ 
    \hline\hline
    $U$      & Fixed Value $[U_\infty]$ 
             & Inlet/ Outlet $[0]$ 
             & Slip / Zero Grad 
             & No Slip \\ \addlinespace
    $p$      & Zero Gradient 
             & Fixed Value $[0]$ 
             & Zero Gradient 
             & Zero Gradient \\ \addlinespace
    $\omega$ & Fixed Value $[0.\overline{6}]$ 
             & Zero Gradient 
             & Zero Gradient 
             & $\omega$-Wall Func. $[5.\overline{3}\cdot 10^{-8}]$ \\ \addlinespace   
    $\nu_t$  & Calculated 
             & Calculated 
             & Calculated 
             & Spalding Wall Func. \\ \addlinespace
    $k$      & Fixed Value $[1.\overline{1}\cdot 10^{-9}]$ 
             & Zero Gradient 
             & Zero Gradient 
             & $k_{qr}$ Wall Function \\ 
    \hline\hline
    \end{tabularx}
\end{table}

\begin{table}
    \centering
    \caption{\label{tab:mesh_convergence} Mesh convergence study for NACA 0012 airfoil at $\alpha=10^\circ$ and $Re=6\times10^6$. Values between brackets refer to the absolute relative error compared to the reference \citep{jespersen2016overflowReference}.}
        \begin{tabular}{c|ll}
        \hline\hline
        Mesh size & $C_D$ ($\varepsilon_{C_D}$,\%) & $C_L$ ($\varepsilon_{C_L}$,\%)\\\hline\hline
        XFOIL   & 0.0097700 (21.87)  & 1.12280 (3.89)\\ \hline
        127286  & 0.0155209 (24.12)  & 1.04594 (3.22)\\ \hline
        508608  & 0.0124938 ~(0.09)  & 1.07599 (0.44)\\ \hline
        544552  & 0.0121113 ~(3.14)  & 1.08158 (0.08)\\ \hline
        808292  & 0.0122211 ~(2.27)  & 1.07856 (0.20)\\ \hline
        1205714 & 0.0123472 ~(1.26)  & 1.07818 (0.23)\\ \hline
        Reference \citep{jespersen2016overflowReference} & 0.0125050 & 1.08075\\ \hline\hline
        \end{tabular}
\end{table}

\subsection{Multi-Fidelity Environment}\label{ss:multifidelity}
The primary objective of the multi-fidelity framework is to leverage the computational efficiency of numerous LF simulations while correcting their accuracy using a sparse set of HF data. Kriging (Gaussian Process Regression, GPR) remains widely used in aeronautics, with strategies ranging from simple difference modeling to complex co-Kriging approaches that learn scaling factors and bias terms simultaneously. In this study, we adopt an LF-informed regression formulation rather than a hierarchical autoregressive co-Kriging model. The multi-fidelity surrogate predicts a target HF aerodynamic quantity, using the corresponding LF output and the airfoil's geometric information. This choice is justified by the performance characteristics of XFOIL: while it effectively captures aerodynamic trends in the linear region, it tends to deviate from RANS predictions in non-linear regimes, particularly regarding lift overestimation near separation. Consequently, the goal is to construct a low-fidelity-informed surrogate that maps $(\boldsymbol{\Theta},\phi_{LF})$ to an estimate of $\phi_{HF}$, providing HF-consistent predictions from LF inputs.

A mapping $f:\phi_{LF}\rightarrow\phi_{HF}$ relying solely on aerodynamic scalars is insufficient. Due to the non-uniqueness in the mapping from LF to HF aerodynamic coefficients, distinct airfoil shapes may yield identical LF performance metrics while exhibiting different HF behavior. To resolve this ambiguity, the geometric parametrization vector $\boldsymbol{\Theta}$ is incorporated into the feature space. In the present implementation, the angle of attack is treated as an operating-condition index rather than as a continuous input to a single global surrogate. For each operating condition $c$, an independent LF-informed GPR model is trained with input

\begin{equation}
\mathbf{x}_c(\boldsymbol{\Theta}) =
\left[
\boldsymbol{\Theta},\phi_{LF,c}(\boldsymbol{\Theta})
\right],
\end{equation}
where $\phi_{LF,c}(\boldsymbol{\Theta})$ is the corresponding XFOIL prediction. The corrected HF estimator is therefore written as
\begin{equation}
\hat{\phi}_{HF,c}(\boldsymbol{\Theta})
=
f_{\mathrm{GP},c}\left(\mathbf{x}_c(\boldsymbol{\Theta})\right)
=
f_{\mathrm{GP},c}\left(\boldsymbol{\Theta},\phi_{LF,c}(\boldsymbol{\Theta})\right)
\end{equation}
where $f_{\mathrm{GP},c}$ denotes the condition-specific Gaussian-process transfer model. Thus, $\boldsymbol{\Theta}$ and $\phi_{LF,c}$ are the actual regression inputs, while $\phi_{HF,c}$ is the supervised target used for training.

To accurately capture the underlying physics while managing the inherent noise of the low-fidelity solver, a composite kernel is employed. The covariance is modeled as a scaled Radial Basis Function (RBF) kernel, which decouples the signal variance (amplitude) from the length scale (correlation). Furthermore, a white noise kernel is added to the covariance matrix, which captures effective observation/modeling noise in the LF-informed input–HF-output mapping, including numerical scatter and convergence artifacts. The resulting kernel function $\psi(x, x')$ is defined as:
\begin{equation}
    \psi(x, x') = \sigma_f^2 \cdot \exp\left(-\frac{||x - x'||^2}{2l^2}\right) + \sigma_n^2 \delta_{x, x'}
\end{equation}
where $\sigma_f^2$ is the signal variance (amplitude), $l$ is the length-scale parameter governing the smoothness of the correlation, $\sigma_n^2$ represents the noise variance, and $\delta_{x, x'}$ is the Kronecker delta function ($1$ if $x=x'$, $0$ otherwise).

For numerical conditioning and reproducibility, all GPR models are trained using standardized inputs and outputs. For each operating condition ($c$), the training inputs $\mathbf{x}_c$ are standardized component-wise using the mean and standard deviation of the initial condition-specific training set $\mathcal{D}^{init}_{c,g}$. The high-fidelity target $\phi_{HF,c}$ is also standardized before training. After prediction, the posterior mean and standard deviation are transformed back to physical aerodynamic units before computing $CV_c$, updating the objective function, or reporting validation errors. Therefore, the normalization used internally for GPR training is separate from the objective-function normalization in Eq.~\eqref{eq:cost_function}. The GPR hyperparameters are optimized by maximizing the log-marginal likelihood using L-BFGS-B (through python's \textit{scikit-learn} library) with 1 independent optimizer restart. The kernel bounds are set to $\sigma_f^2\in[10^{-4},10^4]$, $l\in[10^{-4},10^4]$, and $\sigma_n^2\in[10^{-5},10^1]$. These bounds were kept fixed for all operating conditions and all online retraining steps.

A key element of the proposed framework is the integration of an active learning strategy driven by uncertainty quantification. By exploiting the probabilistic nature of Gaussian Processes, the model provides not only a point prediction but also an estimate of the epistemic uncertainty (prediction variance) at any location in the design space. This capability is leveraged to automatically identify and correct unreliable predictions during the optimization process. A normalized uncertainty metric, the Coefficient of Variation ($CV_c$) is defined for each specific flight condition $c$ as:
\begin{equation}
CV_c(\boldsymbol{\Theta})=
\frac{\sigma_{c}(\mathbf{x}_c(\boldsymbol{\Theta}))}
{|\hat{\phi}_{HF,c}(\boldsymbol{\Theta})|}.
\end{equation}

where $\sigma_c$ is the posterior standard deviation and $\hat{\phi}_{HF,c}$ is the posterior mean prediction of the high-fidelity response.
The normalization produces a dimensionless indicator that can be compared across operating conditions and aerodynamic quantities with different numerical scales. If $CV_c(\boldsymbol{\Theta})>\kappa_c$, the candidate is considered insufficiently certain for surrogate-only evaluation and is escalated to high-fidelity simulation for condition $c$. 
Conversely, if $CV_c(\boldsymbol{\Theta})\leq\kappa_c$, the surrogate prediction is accepted for the objective evaluation. 
Thus, the $CV_c$ threshold is used as a practical fidelity-escalation rule: it does not assume that uncertainty equals error, but prevents highly uncertain surrogate predictions from driving the optimization without RANS validation.

It should be emphasized that $CV_c$ is not intended as a pointwise estimator of surrogate bias or prediction error. Under a well-specified Gaussian-process model, the posterior variance equals the conditional mean-squared deviation of the latent response from the posterior mean, and related Reproducing Kernel Hilbert Space (RKHS) analyses establish high-probability error bounds proportional to the posterior standard deviation under assumptions on the target regularity, noise, and kernel specification \citep{rasmussen2006gaussian, srinivas2010gaussian}. These results motivate the use of posterior uncertainty as an indicator of how strongly a prediction is supported by the available high-fidelity data. They do not, however, imply that a low posterior variance guarantees a low prediction error when the LF-HF relationship is locally misspecified or the covariance model is poorly calibrated \citep{bachoc2013cross}. Accordingly, $CV_c$ is used here only as a dimensionless fidelity-escalation criterion: candidates with large uncertainty relative to the magnitude of the predicted aerodynamic quantity are considered insufficiently reliable for surrogate-only evolutionary selection and are evaluated at high fidelity. Assimilating such a sample contracts the posterior uncertainty and updates the surrogate mean in that region, but no general claim is made that variance reduction alone guarantees bias reduction. Confident but biased predictions are instead mitigated through the supervised LF-informed transfer mapping, online assimilation and retraining with newly acquired RANS samples, mandatory high-fidelity validation of elite candidates, and synchronized re-evaluation of surrogate-only individuals after each surrogate update.

\begin{figure}
    \centering
    \includegraphics[width=8.cm]{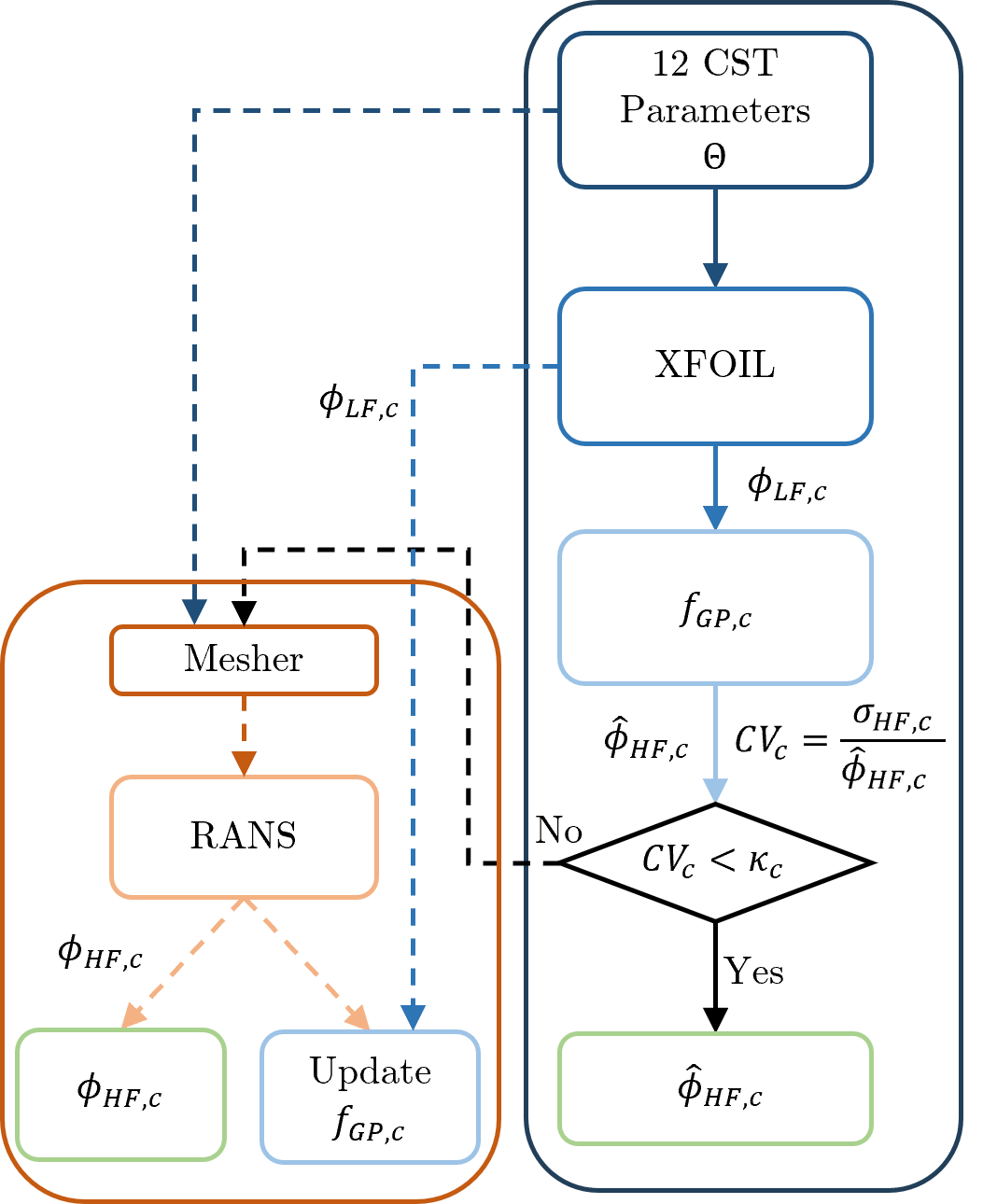}
    \caption{Multi-fidelity evaluation workflow mapping the design vector $\boldsymbol{\Theta}$ through low-fidelity evaluation, Kriging prediction, uncertainty-based refinement, and high-fidelity model updating.}
    \label{fig:03_GPR_multifidelity}
\end{figure}

\begin{algorithm}[t]
    \caption{Condition-wise uncertainty-driven evaluation of an individual.}
    \label{alg:eval_individual}
    \KwRequire{
    Design $\boldsymbol{\Theta}^{(i)}$, operating condition $c$, threshold $\kappa_c$, 
    LF-informed surrogate $f_{\mathrm{GP},c}$, and training set $\mathcal{D}_{c,g}$.
    }
    \KwEnsure{
    $(\hat{\phi}_{HF,c}^{(i)},CV_c^{(i)},HF_c^{(i)},f_{\mathrm{GP},c},\mathcal{D}_{c,g})$.
    }
    Evaluate LF physics $\phi_{LF,c}^{(i)} \leftarrow \mathrm{LF}_c(\boldsymbol{\Theta}^{(i)})$\;
    \eIf{LF solver problem}{
        \textbf{death penalty:} $\hat{\phi}_{HF,c}^{(i)} \leftarrow \emptyset$, 
        $CV_c^{(i)} \leftarrow \infty$, $HF_c^{(i)} \leftarrow 0$\;
    }{
        Construct LF-informed feature vector:
        $\mathbf{x}_c^{(i)} \leftarrow 
        [\boldsymbol{\Theta}^{(i)},\phi_{LF,c}^{(i)}]$\;
        Query surrogate:
        $(\hat{\phi}_{HF,c}^{(i)},\sigma_c^{(i)}) 
        \leftarrow f_{\mathrm{GP},c}(\mathbf{x}_c^{(i)})$\;
        Compute normalized uncertainty:
        $CV_c^{(i)} \leftarrow 
        \sigma_c^{(i)} / |\hat{\phi}_{HF,c}^{(i)}|$\;
        \eIf{$CV_c^{(i)} \leq \kappa_c$}{
            $HF_c^{(i)} \leftarrow 0$\;
        }{
            Run HF solver:
            $\phi_{HF,c}^{(i)} \leftarrow \mathrm{HF}_c(\boldsymbol{\Theta}^{(i)})$\;
            $\hat{\phi}_{HF,c}^{(i)} \leftarrow \phi_{HF,c}^{(i)}$\;
            $HF_c^{(i)} \leftarrow 1$\;
            Update training set:
            $\mathcal{D}_{c,g} \leftarrow 
            \mathcal{D}_{c,g} \cup 
            \{(\mathbf{x}_c^{(i)},\phi_{HF,c}^{(i)})\}$\;
            Re-train condition-wise surrogate:
            $f_{\mathrm{GP},c} \leftarrow \mathrm{Train}(\mathcal{D}_{c,g})$\;
        }
    }
    \Return{
    $(\hat{\phi}_{HF,c}^{(i)},CV_c^{(i)},HF_c^{(i)},f_{\mathrm{GP},c},\mathcal{D}_{c,g})$
    }\;
\end{algorithm}

This condition-wise formulation provides a modular refinement mechanism: each operating point maintains its own LF-HF transfer model and is enriched only when the corresponding uncertainty criterion is violated. Therefore, a candidate does not necessarily require high-fidelity evaluation at all operating conditions, but only at those for which the local surrogate reliability is insufficient. This is useful when the LF-HF discrepancy is strongly condition-dependent, as in the present case. Nevertheless, if the number of operating conditions becomes large, maintaining independent surrogates and thresholds may increase the modeling and algorithmic burden. In such cases, sparse or multi-output Gaussian-process formulations, shared latent representations, or clustering of operating conditions could be used to preserve scalability while retaining the same active-refinement principle.

The calibration of the uncertainty thresholds $\kappa_c$ is performed immediately following the initialization phase. Once the initial surrogate baseline is established using the initial training set, a calibration study is conducted using 1000 randomly sampled individuals evaluated via the low-fidelity solver, different from those used for initialization of the optimizer. These samples are then queried against the multi-fidelity models to generate HF estimates and their associated uncertainty metrics (i.e, $CV_c$). 
To balance computational cost with model robustness, the thresholds $\kappa_c$ are defined at the $90^\text{th}$ percentile of the resulting uncertainty distribution. This criterion limits HF calls to the most uncertain candidates (approximately 10\% under the calibration distribution), providing a pragmatic balance between refinement and cost. Other percentile choices are possible and would trade HF budget against surrogate correction aggressiveness; the $90^\text{th}$ percentile was selected to keep HF usage bounded while still correcting excursions into uncertain regions. This procedure yielded thresholds of $\kappa_{\alpha=2^\circ}=0.05$ and $\kappa_{\alpha=10^\circ}=0.02$, which are held fixed for the remainder of the optimization. 

The marked asymmetry between these two thresholds reflects a structural difference in the LF–HF discrepancy across operating conditions. At cruise ($\alpha=2^\circ$), the target objective is aerodynamic efficiency ($L/D$), which depends critically on accurate drag prediction. Two factors compound the LF–HF mismatch for this quantity. First, XFOIL's viscous–inviscid interaction formulation is inherently limited in its ability to predict drag with the same fidelity as lift, particularly in regimes where pressure recovery and incipient separation influence the drag budget. Second, the two solvers adopt fundamentally different assumptions regarding the boundary-layer state: XFOIL employs an $e^N$ transition model that can predict extended laminar regions and the associated lower skin-friction drag. At cruise, where the drag pressure contribution is not as pronounced, this discrepancy is exacerbated, leading to systematic differences in drag estimation that the surrogate must correct through a more permissive uncertainty threshold. At take-off  ($\alpha=10^\circ$), the strong adverse pressure gradient near the leading edge forces natural transition far upstream, so that both solvers effectively model a predominantly turbulent boundary layer. As a result, the LF–HF agreement for($C_L$) at this condition is already strong, and only a tight threshold ($\kappa_{\alpha=10^\circ}=0.02$) is needed to capture the residual discrepancy. This physical interpretation is consistent with the surrogate validation results presented in \autoref{s:Results}, where the RMSRE reduction achieved by the transfer mapping is substantially larger for cruise efficiency than for take-off lift.

\subsection{Optimization Problem Formulation} \label{ss:Cost_function}
The multi-condition objectives of the study are encapsulated in a single scalar cost function $J$, which the algorithm seeks to minimize. The design goal is twofold: first, to maximize aerodynamic efficiency ($E = L/D$) at the cruise condition ($\alpha=2^\circ$); and second, to maximize the lift coefficient ($C_L$) at the take-off condition ($\alpha=10^\circ$). The latter is critical for enhancing operational safety and reducing reliance on complex high-lift devices. Under cruise conditions, directly modeling $E$ yielded optimization trends comparable to using separate surrogate models for $C_L$ and $C_D$; however, the latter introduces additional complexity (including uncertainty/error propagation through the ratio) and was therefore not retained in the present implementation.

To formulate this multi-objective maximization problem as a single-objective minimization task, the cost function is defined as the inverse of a weighted sum:
\begin{equation}\label{eq:cost_function}
    J = \frac{1}{0.5 \cdot \hat{E}^{\alpha=2^\circ} + 0.5 \cdot \hat{C}_L^{\alpha=10^\circ} + \varepsilon}
\end{equation}
where $\varepsilon$ is a damping constant, and $\hat{E}$ and $\hat{C}_L$ represent the normalized aerodynamic efficiency and lift coefficient, specifically, $\hat{E}^{\alpha=2^\circ} = E^{\alpha=2^\circ}/\overline{E}_0^{\alpha=2^\circ}$ and $\hat{C}_L^{\alpha=10^\circ} = C_L^{\alpha=10^\circ}/\overline{C}_{L,0}^{\alpha=10^\circ}$, where $\overline{(\cdot)}_0$ denotes the mean over the initial population. 
Employing the performance values observed in the initial population for normalization ensures that both metrics possess comparable magnitudes, which enables the use of equal weighting ($w=0.5$) as a balanced two-condition demonstration. The selected weights are therefore not intended to be universal, but rather to define a neutral scalarization for the present cruise/take-off test case. In practical applications involving additional operating points, the weights should be prescribed according to mission-level priorities, such as the relative importance of cruise efficiency, climb performance, take-off capability, or landing constraints. Alternatively, the same active multi-fidelity refinement strategy could be embedded within a multi-objective optimizer, avoiding the need to prescribe a single scalarization a priori. On the other hand, the damping constant $\varepsilon$ regularizes the reciprocal formulation by limiting fitness amplification when the weighted sum becomes small due to early-iteration surrogate errors or solver outliers. In the present study, $\varepsilon=10$ was selected pragmatically to moderate selection pressure while preserving the relative ranking of candidates, since the normalized terms remain $\mathcal{O}(1)$ over the explored design space.

The optimization is initialized using Latin Hypercube Sampling (LHS) to ensure a space-filling distribution of the initial population. A total of 100 candidate geometries were sampled and subjected to both low- and high-fidelity simulations. Of these, 83 converged successfully in both solvers, with the non-converging cases attributed to extreme geometric complexities, mesh-generation, or solver-convergence issues. These 83 valid individuals constitute the initial population used to train the baseline surrogate models.

\subsection{Optimization Framework}\label{ss:Optimization_framework}
The present study addresses a multi-condition airfoil design problem over a 12-parameter CST space, without warm-starting from an existing baseline beyond the parameter bounds and geometric feasibility constraints. A hybrid genetic algorithm, HyGO \citep{robledo2025hygo}, is employed as the global optimizer for this task. Genetic optimization provides global-search capability, which supports exploration of weakly constrained design spaces, but it can require a large number of evaluations to converge. HyGO addresses this limitation by hybridizing a standard GA with a local-search Downhill Simplex algorithm (DSM) \citep{Nelder1965DSM} to improve exploitation and accelerate convergence. In HyGO, the CST parameters are encoded using a fixed-length binary representation ($N_b^{f}$ bits per parameter) and decoded to continuous values within the bounds in \autoref{tab:CST_domain}. 

The hyperparameters governing the optimization are reported in \autoref{tab:hygo_hyperparameters}. The HyGO run is limited to a maximum of $N_g=15$ generations, with an early-stopping criterion triggered when the best objective value does not improve for three consecutive generations. To promote broad exploration of the 12-parameter CST space in the early stages, the exploration population size is set to $N_{explor}^{2-4}=75$ for Generations 2–4 and reduced to $N_{explor}^{5-15}=50$ from Generation 5 onward to concentrate computational effort on refinement. The evolutionary selection follows the tournament procedure described in \citet{robledo2025hygo}, with crossover and mutation applied with probabilities $P_c=0.55$ and $P_m=0.45$ and a bit-flip probability $p_m=0.05$. The hybridization stage performs local-search exploitation on $N_{\mathrm{exploit}}=15$ candidates per generation using the Downhill Simplex method. These settings were selected pragmatically based on prior HyGO performance on comparable engineering optimization problems and the desired exploration–exploitation trade-off under the available evaluation budget. For further details on the algorithm and its hyperparameters, refer to \citet{robledo2025hygo}.

\begin{table}
\centering
  \caption{Optimization hyperparameters for HyGO}
  \label{tab:hygo_hyperparameters}
  \resizebox{8cm}{!}{%
      \begin{tabular}{ccl}
        \hline\hline
        Name & Value & Description\\
        \midrule
        \midrule
         $N_b^{f}$ & 5 & Bit number per parameter\\
         $N_g$ & 15 & Total number of generations\\
         $N_{init}$ & 83 & Initialization size\\
         $N_{explor}^{2-4}$ & 75 & Gen. 2 - 4 Exploration size\\
         $N_{explor}^{5-15}$ & 50 & Gen. 5 - 15 Exploration size\\
         $N_T$ & 7 & Tournament size\\
         $p_T$ & 1 & Tournament selection probability \\
         $P_c$ & 0.55& Crossover probability\\
         $P_m$ & 0.45& Mutation probability\\
         $p_m$ & 0.05& Mutation Bit flip Probability\\
         $N_e$ & 3   & Elitism individuals\\
         $d_e$ & $0.05\cdot\mathcal{L}_\Theta$   & Elitism minimum distance\\
         $N_{exploit}$ & 15 & Exploitation offspring\\
         $N_s$ & 13 & Simplex Size\\
        \hline\hline
    \end{tabular}}
\end{table}

The present work modifies the original HyGO workflow in three aspects required by the active multi-fidelity setting. First, each candidate is evaluated through condition-wise LF-informed surrogate models, with high-fidelity escalation triggered only when the corresponding uncertainty threshold is exceeded. Second, elite candidates are systematically validated with the high-fidelity solver before being propagated to the next generation. Third, whenever new high-fidelity samples are assimilated and the surrogate is retrained, surrogate-only candidates are re-evaluated to synchronize their costs with the current surrogate state.

\begin{algorithm*}[h!]
    \caption{Modified evolution strategy with uncertainty-triggered fidelity selection and synchronized elitism. The condition-wise individual evaluation is described in \autoref{alg:eval_individual}.}
    \label{alg:elitism_procedure}

    \KwRequire{Generation $g=1$, maximum generations $N_g$, elitism size $N_e$, 
    exploration/exploitation sizes $N_{\mathrm{explore}},N_{\mathrm{exploit}}$, 
    condition-wise thresholds $\{\kappa_c\}$, datasets $\{\mathcal{D}_{c,g}\}$, 
    and surrogates $\{f_{\mathrm{GP},c}\}$.}
    \KwEnsure{Best design $\boldsymbol{\Theta}^\star$.}

    \While{not converged \textbf{and} $g < N_g$}{

      \BlankLine
      \tcp*[l]{\hfill --- Exploration --- \hfill}
      Generate $N_{\mathrm{explore}}$ offspring from the elite set using HyGO genetic operators\;
      \ForEach{exploration candidate $\boldsymbol{\Theta}^{(i)}$}{
            \ForEach{condition $c\in\mathcal{C}$}{
          Evaluate candidate with \autoref{alg:eval_individual};
          update $\{\mathcal{D}_{c,g}\}$ and $\{f_{\mathrm{GP},c}\}$ if HF is triggered\;
          }
      }
      Synchronize surrogate-only candidates using the updated surrogates\;
      Sort exploration population by synchronized cost\;

      \BlankLine
      \tcp*[l]{\hfill --- Exploitation --- \hfill}
      $n_{\mathrm{exploit}}\leftarrow 0$\;
      \While{$n_{\mathrm{exploit}}<N_{\mathrm{exploit}}$}{
          Generate $N_c$ candidates using DSM operations\;
          \ForEach{DSM-generated candidate $\boldsymbol{\Theta}^{(j)}$}{
          \ForEach{condition $c\in\mathcal{C}$}{
              Evaluate candidate with \autoref{alg:eval_individual};
              update $\{\mathcal{D}_{c,g}\}$ and $\{f_{\mathrm{GP},c}\}$ if HF is triggered\;
              }
          }
          $n_{\mathrm{exploit}}\leftarrow n_{\mathrm{exploit}}+N_c$\;
      }
      Sort the full population by cost\;

      \BlankLine
      \tcp*[l]{\hfill --- HF-validated elitism and synchronization --- \hfill}
      Select elite set $\mathcal{E}_g$ using cost and minimum-distance criterion\;
      \ForEach{elite candidate $e\in\mathcal{E}_g$}{
      \ForEach{condition $c\in\mathcal{C}$}{
          Run missing HF evaluations for all required conditions\;
          update the corresponding $\mathcal{D}_{c,g}$ and retrain $f_{\mathrm{GP},c}$\;
          assign HF-validated cost $J^{(e)}$\;
          }
      }

      Re-evaluate all surrogate-only candidates with the updated surrogates\;
      Assign $J=\infty$ to candidates that violate $\exists c \in \mathcal{C} : CV_c(\Theta) > \kappa_c$ after synchronization\;
      Sort population by synchronized costs\;
      $g\leftarrow g+1$\;
    }
    \Return{$\boldsymbol{\Theta}^\star=\arg\min_{\boldsymbol{\Theta}}J(\boldsymbol{\Theta})$}\;
\end{algorithm*}

The modification can be described in terms of the optimization state at generation $g$. Let $\mathcal{P}_g$ denote the population generated by HyGO, $\mathcal{E}_g\subset\mathcal{P}_g$ the set of elite candidates, and $\mathcal{D}_{c,g}$ the high-fidelity training set associated with operating condition $c$. Each condition has an associated surrogate state $\mathcal{M}_{c,g}=\{f_{GP,c},\kappa_c\}$, where $f_{GP,c}$ is trained on $\mathcal{D}_{c,g}$ and $\kappa_c$ is the corresponding uncertainty threshold.

During the generation cycle, candidates are first evaluated using the condition-wise surrogate models, obtaining the posterior prediction $\hat{\phi}_{HF,c}$ and the normalized uncertainty indicator $CV_c$. If $CV_{c}(\boldsymbol{\Theta})\leq\kappa_c$, the cost contribution associated with condition $c$ is computed from the surrogate prediction. If $CV_{c}(\boldsymbol{\Theta})>\kappa_c$, a high-fidelity evaluation is performed for the corresponding condition, the pair $(\mathbf{x}_c(\boldsymbol{\Theta}),\phi_{HF,c}(\boldsymbol{\Theta}))$ is appended to $\mathcal{D}_{c,g}$, and the surrogate model $f_{GP,c}$ is retrained. The scalar objective $J$ is then computed from the set of condition-wise predictions or high-fidelity values available for the candidate.

Once the population has been evaluated, the elite set $\mathcal{E}_g$ is selected using both cost and geometric diversity. The best candidate is always retained, while the remaining $N_e-1$ elite candidates are accepted only if they satisfy a minimum Euclidean distance
\begin{equation}
d_e = 0.05\,\mathcal{L}_\Theta
\end{equation}
from all previously selected elites, where $\mathcal{L}_\Theta$ is the mean euclidean distance between existing elite and candidate CST parameters. This diversity-aware elitism prevents the elite set from collapsing into nearly identical candidates, a common outcome when local DSM exploitation is combined with evolutionary selection. Since elite candidates directly determine the genetic material propagated to the next generation, all selected elites are mandatorily validated with the high-fidelity solver whenever a high-fidelity value is not already available for the required operating conditions.

The high-fidelity evaluations acquired through uncertainty triggering or elite validation modify the surrogate state within the same generation. Therefore, candidates previously evaluated only by the surrogate may have stale cost values. A synchronization step is consequently applied: all surrogate-only candidates are re-evaluated with the updated surrogate models, and their costs are recomputed. If a candidate becomes uncertain after synchronization, i.e., $CV_{c}>\kappa_c$, it is penalized rather than recursively sent to high-fidelity evaluation. This avoids unbounded inner loops of simulation, surrogate retraining, and re-evaluation, while keeping the per-generation high-fidelity budget controlled, as well as avoiding uncertain information propagation through the generations. The population is then ordered according to the synchronized costs and passed to the next HyGO generation. Algorithm~\ref{alg:elitism_procedure} summarizes this modified evolution strategy, whereas the standard crossover, mutation, tournament selection, and Downhill Simplex operations follow the original HyGO formulation in~\citep{robledo2025hygo}.

The proposed workflow assumes that both the low-fidelity (XFOIL) and high-fidelity (RANS) solvers return converged, physically meaningful outputs. In practice, solver instability and the wide design space enabled by the CST parametrization can lead to non-convergence or invalid solutions. At the low-fidelity level, XFOIL provides explicit warning flags upon failure; whenever such a warning is triggered, the corresponding individual is assigned a death penalty ($J=\infty$) before the uncertainty evaluation, ensuring that it is immediately discarded by the optimizer. 
This choice introduces a bias against regions where the LF model fails, but it is acceptable here because the target operating conditions ($\alpha=2^\circ$ cruise and $\alpha=10^\circ$ take-off at $\mathrm{Re}=6\times10^6$) are not intended to explore deep-stall behavior; designs that systematically trigger XFOIL failure typically correspond to strongly separated regimes outside the validity range of the LF solver and outside the intended mission envelope. The alternative of directly evaluating the HF in such cases introduces an uncontrolled triggering mechanism that may easily overflow the simulation budget without providing meaningful information.
An analogous screening is applied to high-fidelity evaluations. A high-fidelity result is considered admissible only if all of the following conditions are satisfied: (i) the simulation reaches the prescribed maximum number of iterations without divergence or the residual criterion is met; (ii) over the last 2,000 iteration steps, the standard deviations of $C_L$ and $C_D$, normalized by their respective means, are below 0.05; and (iii) both $C_L$ and $C_D$ remain positive to exclude non-physical solver outputs. This screening procedure increases robustness by ensuring that only converged, physically consistent high-fidelity outputs are used to evaluate the cost function and update the surrogate models.

Finally, the framework incorporates a geometric validity check based on the constraints defined in \autoref{ss:parametrization}. Leveraging HyGO's feasibility handling mechanism, individuals flagged as invalid are first subject to regeneration attempts via genetic operators. If a candidate fails to meet the validity criteria after a maximum number of regeneration attempts (1000 in this work), it is discarded. This strategy enhances computational efficiency by preventing the allocation of simulation resources to \textit{a priori} infeasible designs, effectively confining the search to the productive regions of the parametric space.

\section{Results}\label{s:Results}
\begin{figure}
    \centering
    \includegraphics[width=8.cm]{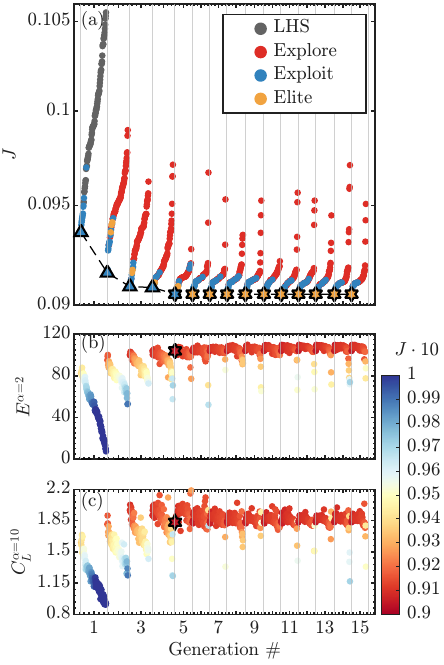}
    \caption[]{Optimization convergence and objective evolution over 15 generations: cost history, generation operators, cruise efficiency, take-off lift, and overall best design.}
    \label{fig:05_J_evolution}
\end{figure}

The optimization campaign evaluated a total of 1042 individuals across 15 generations. The evolution of the cost function is depicted in \autoref{fig:05_J_evolution}. The best individual was effectively achieved by generation 5 (marked by \tikz\node[star, fill=blueelite, star points=6, draw=black, star point ratio=2.5, inner sep=1.5pt] {};), followed by a plateau in performance until generation 15, highlighting convergence. The results demonstrate that the algorithm successfully refines the initial population generated via Latin Hypercube Sampling. Notably, the hybrid exploitation mechanism (DSM) played a critical role in the early stages, producing several high-performing individuals in the first two generations that significantly accelerated convergence toward the best-found design.
\begin{table} 
    \caption{\label{tab:Optimization_resultsm} Optimization results for the 12-parameter CST representation, including the aerodynamic performance metrics. Results for the best individual after the first generation and the overall best reached after the fifth generation.}
    \centering
    \begin{tabular}{l|cc}
        \hline\hline
        Variable & Gen. 1 & Gen. 5--15 \\ 
        \hline\hline
        $C_L^{\alpha=10^\circ}$ & 1.518 & 1.833 ($20.75\%\; \uparrow$)\\
        $E^{\alpha=2^\circ}$    & 79    & 111.43 ($41.05\%\; \uparrow$)\\ 
        \hline
        $\theta_1$    &  0.43961 &  0.27546\\
        $\theta_2$    &  0.29592 &  0.12594\\
        $\theta_3$    &  0.66546 &  0.53639\\
        $\theta_4$    &  0.05585 &  0.17579\\
        $\theta_5$    &  0.34700 &  0.34700\\
        $\theta_6$    &  0.09677 & -0.16129\\
        $\theta_7$    & -0.11924 & -0.16335\\
        $\theta_8$    & -0.27301 & -0.16790\\
        $\theta_9$    & -0.00118 &  0.10869\\
        $\theta_{10}$ &  0.04885 &  0.40348\\
        $\theta_{11}$ & -0.07258 &  0.15619\\
        $\theta_{12}$ &  0.08158 & -0.03860\\ 
        \hline\hline
    \end{tabular}
\end{table}

The quantitative outcomes of the optimization campaign are summarized in \autoref{tab:Optimization_resultsm}. A direct comparison between the best individual from the first generation (which notably already benefited from local search exploitation) and the final optimized geometry highlights the substantial performance improvements achievable through the optimization process. The method yielded a 20.75\% increase in $C_L^{\alpha=10^\circ}$ and a 41.05\% improvement in $E^{\alpha=2^\circ}$. Physically, the final take-off lift $C_L^{\alpha=10^\circ}=1.833$ is high for a single-element airfoil at the specified operating condition. Such lift levels are typically associated with strongly cambered geometries and often accompanied by a significant drag penalty due to increased pressure drag and thicker boundary layers. In the present case, however, the configuration also exhibits a high aerodynamic efficiency $E^{\alpha=2^\circ}=111.43$ despite the pronounced camber of the optimized airfoil. This indicates that the optimization procedure has identified a pressure distribution that enhances lift generation while maintaining a comparatively favorable recovery in the aft region, thereby limiting the drag increase that would normally accompany such high lift coefficients. Regarding geometry representation, the final design reveals that the CST parameters converged to intermediate values, well within the user-defined bounds. This interior solution suggests that the parametric envelope was sufficiently broad. 

\begin{figure}[t]
    \centering
    \includegraphics[width=8.cm]{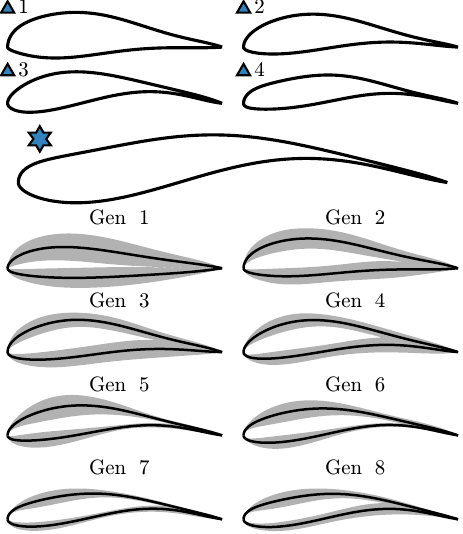}
    \caption[]{Geometric evolution of optimized airfoils, showing generation-best profiles, the final optimum, and population envelopes with mean geometries over Generations 1--8.}
    \label{fig:05_J_evolution_airfoils}
\end{figure}

The optimized geometric topology is depicted in \autoref{fig:05_J_evolution_airfoils} (marked by the blue star \tikz\node[star, fill=blueelite, star points=6, draw=black, star point ratio=2.5, inner sep=1.5pt] {};). The resulting shape is non-standard within the sampled envelope, characterized by pronounced aft-loading and extreme camber peaking near $x/c \approx 0.5$. This aft-camber distribution is efficient for lift generation but introduces a geometric inflection on the lower surface. Consequently, the aft region of the pressure side is subjected to a steep adverse pressure gradient, with risk of flow separation at low angles of attack (such as the cruise condition, $\alpha=2^\circ$). Furthermore, the thickness distribution is noticeably decoupled from the camber; the maximum thickness is concentrated in the forward characteristic length, while the curvature drives the aft section. Notably, the upper surface exhibits a flattened, quasi-linear region immediately downstream of the leading edge. This geometry aligns the surface almost parallel to the local inflow at $\alpha=10^\circ$, effectively delaying the suction peak, broadening the suction distribution, and delaying separation onset.

The evolution of the specific aerodynamic objectives is detailed in \autoref{fig:05_J_evolution}. Rapid improvements in both metrics are observed in the initial generations. Interestingly, a subset of early individuals (corresponding to generations 2-3) exhibits higher lift coefficients at take-off than the final design. Analysis of the representative geometric profiles (bottom of \autoref{fig:05_J_evolution_airfoils}) reveals that this high lift was primarily driven by increased airfoil thickness. However, excessive thickness incurs a substantial drag penalty, detrimental to the cruise efficiency objective. Consequently, later generations move toward a geometric compromise, balancing reduced thickness with optimized camber to maximize the composite objective function. This trend is visually evident in the optimized profiles of the first five generations (top-left inset), where the geometry progressively evolves from thick, high-lift shapes to thinner, more cambered designs that sustain lift while minimizing drag.

\begin{figure}[t]
    \centering
    \includegraphics[width=8.cm]{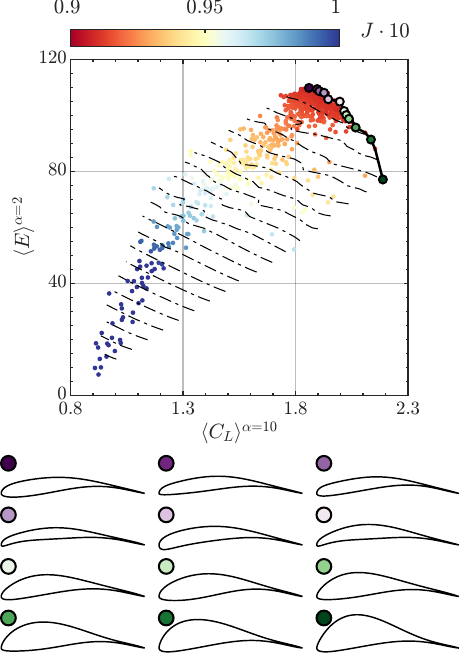}
    \caption{Composite objective space showing cruise efficiency, take-off lift, cost-colored samples, Pareto-optimal front, and corresponding airfoil geometries along the front.}
    \label{fig:06_pareto}
\end{figure}

The strong alignment between the two aerodynamic objectives over much of the sampled design space is corroborated in \autoref{fig:06_pareto}, which shows an approximately linear trend across the majority of evaluated candidates. The dashed curves \lcap{.-}{black} represent iso-contours of the composite cost $J$ in the ($C_L^{\alpha=10^\circ},E^{\alpha=2^\circ}$) plane, obtained by evaluating $J$ on a grid of objective values; they are shown as a visual reference for the direction of decreasing cost. Over most of the explored region, increases in take-off lift ($C_L^{\alpha=10^\circ}$) tend to be accompanied by increases in cruise efficiency ($E^{\alpha=2^\circ}$), indicating that, within the imposed CST bounds and feasibility constraints, early improvements are only weakly conflicting.

However, this trend breaks down in the upper-right quadrant, where the highest-performing individuals reside, and a distinct Pareto front becomes apparent (marked by a solid black line \lcap{-}{black}). This region highlights the local trade-off between further lift gains at take-off and efficiency retention at cruise, consistent with increasing drag sensitivity at high-lift geometries. The bottom half of \autoref{fig:06_pareto} displays 12 of the 23 Pareto-optimal airfoils and shows smooth, gradual shape transitions along the front.  In line with the trends observed in \autoref{fig:05_J_evolution_airfoils}, the high-$C_L$ end of the front (green markers \tikz\node[circle, fill=greenregion, draw=black, inner sep=2.1pt] {};) is dominated by thicker profiles, whereas intermediate Pareto solutions (purple markers  \tikz\node[circle, fill=purpleregion, draw=black, inner sep=2.1pt] {};) shift toward increased camber and reduced leading-edge radius.

Notably, the evaluated set does not contain solutions combining high $E$ with low $C_L$, or vice-versa. This absence is consistent with the fixed-weight scalarization and feasibility constraints: designs with reduced take-off lift are strongly disfavored by the composite objective, so the search concentrates on regions where efficiency gains are achieved without sacrificing $C_L^{\alpha=10^\circ}$. In addition, maximizing $E^{\alpha=2^\circ}=L/D$ alone does not necessarily promote very low-lift geometries, since $E$ depends on the coupled behavior of lift and drag rather than on drag minimization in isolation.

\begin{figure}[t]
    \centering
    \includegraphics[width=8.cm]{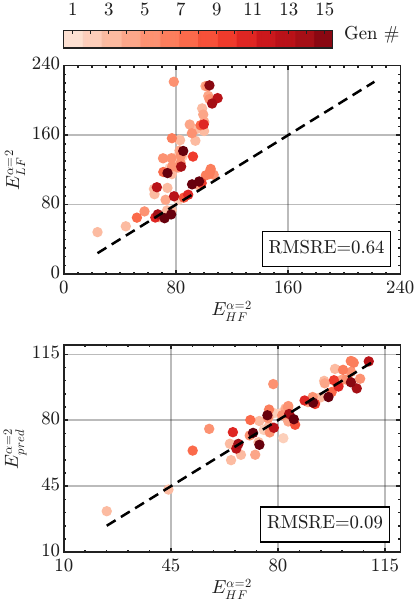}
    \caption{Cruise-condition surrogate validation comparing high-fidelity references with XFOIL outputs and pre-update multi-fidelity predictions, with samples colored by acquisition generation.}
    \label{fig:07_rans_vs_pred_aoa2}
\end{figure}

\begin{figure}[t]
    \centering
    \includegraphics[width=8.cm]{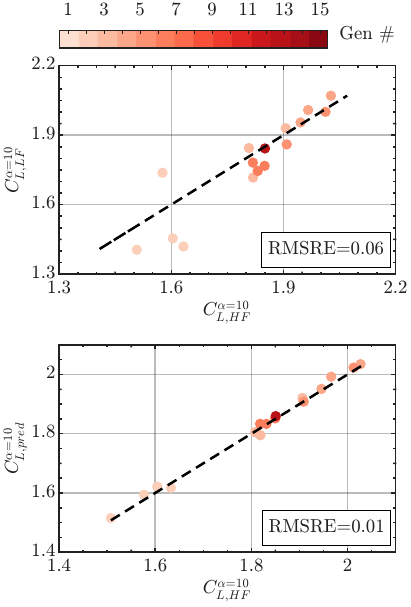}
    \caption{Take-off-condition surrogate validation comparing high-fidelity references with XFOIL outputs and pre-update multi-fidelity predictions, with samples colored by acquisition generation.}
    \label{fig:07_rans_vs_pred_aoa10}
\end{figure}

\autoref{fig:07_rans_vs_pred_aoa2} and \autoref{fig:07_rans_vs_pred_aoa10} assess the predictive performance of the proposed LF-informed surrogate. In both figures, the HF reference values are compared against the LF outputs and the surrogate predictions (identified by subindex $pred$) evaluated immediately before assimilating the corresponding HF samples into the model update. This pre-update evaluation avoids optimistic ``in-sample'' assessment and isolates the effect of the transfer mapping on newly acquired HF points. 

For the cruise condition ($\alpha=2^\circ$) shown in \autoref{fig:07_rans_vs_pred_aoa2}, XFOIL exhibits a systematic bias in aerodynamic efficiency, yielding LF values that frequently exceed the HF range encountered in the optimization. This behavior is consistent with the sensitivity of $E$ to drag prediction and with the known limitations of viscous–inviscid approaches in regimes where drag is influenced by separation-related mechanisms. The Kriging-based transfer mapping substantially reduces this bias and the associated scatter, bringing the predictions closer to the identity line. Quantitatively, the root mean squared relative error (RMSRE), computed on each newly acquired HF point before its assimilation and then aggregated across generations, decreases from $0.64$ (LF) to $0.09$ (pred),  i.e., an improvement by approximately a factor of seven. Additionally, the generation-based color coding indicates that later acquisitions tend to concentrate closer to the identity line, which is consistent with progressive local correction of the LF–HF mapping along the optimizer trajectory as additional HF samples are assimilated.

For the take-off condition ($\alpha=10^\circ$) in \autoref{fig:07_rans_vs_pred_aoa10}, the LF-HF agreement for $C_L$ is already strong, and the transfer surrogate primarily reduces residual dispersion. The RMSRE decreases from $0.06$ (LF) to $0.01$ (pred), indicating that only a modest correction is required for $C_L$ at this condition. Consistent with this, most late-generation HF evaluations are expected to originate from the mandatory HF validation of elite individuals, whereas uncertainty-triggered escalation is comparatively limited for the $C_L$ model.

These results also clarify the condition-dependent role of the multi-fidelity correction. The LF model is not uniformly inaccurate across all objectives: at the take-off condition, $\alpha=10^\circ$, the LF prediction of $C_L$ is already strongly aligned with the HF response, so only limited correction is required. In contrast, at the cruise condition, $\alpha=2^\circ$, the efficiency metric $E=L/D$ exhibits a substantially larger LF--HF discrepancy because small drag-prediction errors are amplified through the lift-to-drag ratio. This is precisely the regime where online HF enrichment becomes most important, since an LF-only or static-surrogate optimization may favor candidates with unrealistically high predicted efficiencies. The active multi-fidelity updates progressively correct this bias along the optimization trajectory and prevent the evolutionary search from being driven by misleading LF trends.

\begin{figure}[t]
    \centering
    \includegraphics[width=8.cm]{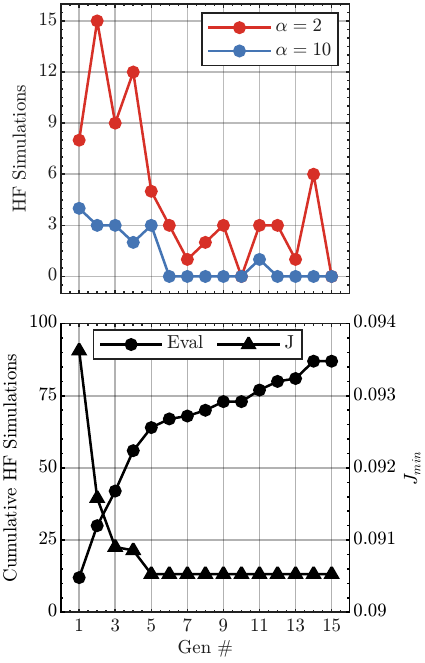}
    \caption{High-fidelity budget per generation. Top: Number of HF simulations executed at each flight condition as a function of generation. Bottom: Cumulative number of HF simulations for both conditions and the best cost per generation.}
    \label{fig:08_gens}
\end{figure}

This behavior in later generations is further quantified by the number of HF simulations per generation, presented in \autoref{fig:08_gens}. Although the best design was first identified by Generation 5 and the campaign satisfies the HyGO convergence criterion by Generation 8 (no improvement over three consecutive generations), the run was intentionally continued to Generation 15. This extended phase isolates the effect of online surrogate refinement and illustrates how uncertainty-triggered HF escalation diminishes as the LF-informed transfer models become better calibrated in the regions visited by the optimizer, illustrated in the bottom panel of \autoref{fig:08_gens} by the correlation between the cumulative simulations and the minimal cost evolution. 

\begin{figure}[t] 
    \centering 
    \includegraphics[width=8.cm]{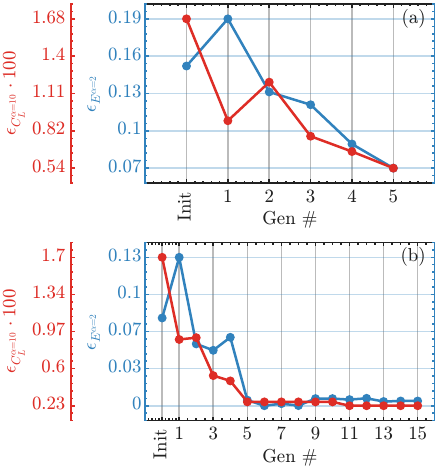}
    \caption[]{Online GPR convergence under high-fidelity enrichment, showing future-trajectory prediction errors and final-design landmark errors for cruise efficiency and take-off lift.} \label{fig:13_GPR_error_evo}
\end{figure}

To further assess the effect of the online updates on the surrogate itself, a convergence analysis was performed. For each generation, the corresponding condition-wise GPR state was reconstructed and evaluated on RANS-validated candidates relevant to the optimization trajectory. The relative error for a target quantity $q\in\{E^{\alpha=2^\circ},C_L^{\alpha=10^\circ}\}$ was computed as 
\begin{equation} 
\mathcal{E}_{q}^{(g)}= \frac{1}{|\mathcal{T}|} \sum_{\boldsymbol{\Theta}\in\mathcal{T}} \left| \frac{\hat{q}^{(g)}(\boldsymbol{\Theta})-q_{HF}(\boldsymbol{\Theta})} {q_{HF}(\boldsymbol{\Theta})} \right|,
\end{equation} 
where $\hat{q}^{(g)}$ denotes the prediction of the GPR model available after generation $g$, and $\mathcal{T}$ is the evaluation set. Two complementary tests are shown in \autoref{fig:13_GPR_error_evo}. In panel (a), $\mathcal{T}$ is defined as the set of RANS-evaluated candidates acquired after the best design was first identified, i.e., candidates from Generations 6--15. The GPR snapshots from initialization to Generation 5 are evaluated on this fixed future-trajectory set. The decreasing errors indicate that the online updates improve the surrogate in the region subsequently explored by the optimizer, rather than only improving a global random validation metric. In panel (b), the final best design is used as a landmark point to assess how the online updates correct the surrogate prediction in the region ultimately selected by the optimizer (excluded from the GPR training for this analysis). This second analysis should be interpreted as a trajectory-local correction diagnostic, since the error naturally decreases once high-fidelity information from the selected region has been assimilated into the model. Together, these results support the role of online GPR updates in maintaining surrogate reliability along the optimization path, improving confidence in the found minima. This analysis should not be interpreted as a theoretical proof that posterior uncertainty is equivalent to prediction error; rather, it provides an a posteriori verification that the combined update strategy reduces surrogate error in the optimization-relevant region for the present problem.

To further isolate the role of the online refinement mechanism, a static-surrogate ablation was performed and is reported in ~\autoref{anex:static_GPR}. The same initial high-fidelity training set, LF-informed GPR formulation, scalar objective, and HyGO hyperparameters were retained, but the surrogate was kept fixed after initialization. Consequently, uncertainty-triggered enrichment, mandatory high-fidelity elite validation, and population synchronization were disabled. This ablation is not intended as an exhaustive comparison against fixed-budget enrichment, expected-improvement infill, adaptive Kriging, or other active-learning policies. Rather, it is a controlled test of the specific failure mode targeted in this work: surrogate overexploitation during evolutionary selection when no online high-fidelity correction is available. Although the static surrogate produced an apparently lower predicted cost, post hoc RANS validation revealed a surrogate-induced false optimum: the predicted values $E^{\alpha=2^\circ}=117.24$ and $C_L^{\alpha=10^\circ}=2.96$ decreased to $E^{\alpha=2^\circ}=20.92$ and $C_L^{\alpha=10^\circ}=2.235$ under RANS evaluation. This result shows that, in the present open CST design space, the proposed online refinement is not only a cost-reduction mechanism but also a safeguard against propagating candidates whose predicted performance is unsupported by high-fidelity data.

The resulting HF budget exhibits a marked asymmetry between operating conditions. For the cruise condition ($\alpha=2^\circ$), a comparatively large number of HF calls is required in early generations (e.g., in Generation 1, 8 of the 20 exploitation candidates exceed the uncertainty threshold). Exploration candidates in this phase do not trigger additional HF evaluations, since they are drawn from the initial LHS set already evaluated with RANS. The sustained need for HF updates at $\alpha=2^\circ$ is consistent with the greater difficulty of transferring efficiency $E=L/D$, which is highly sensitive to drag prediction and therefore amplifies LF–HF discrepancies. In contrast, for the take-off condition ($\alpha=10^\circ$), the LF–HF agreement for $C_L$ is strong, and uncertainty-triggered escalation rapidly vanishes.
\begin{figure}[t]
    \centering
    \includegraphics[width=8.cm]{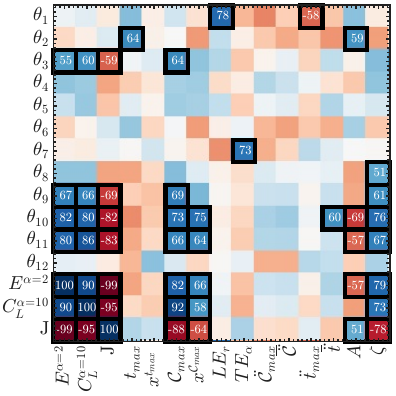}
    \caption{Pearson correlations between CST parameters, geometric descriptors, and aerodynamic quantities, with boxed entries identifying dominant linear associations for interpretability.}
    \label{fig:09_corr}
\end{figure}

Overall, the campaign executed 154 and 99 HF evaluations at $\alpha=2^\circ$ and $\alpha=10^\circ$, respectively, including the initial set of 83 geometries evaluated directly at high fidelity. Since the complete optimization campaign involved 1042 candidate evaluations, an equivalent all-HF reference campaign using the same RANS workflow at both operating conditions would require $1042\times2=2084$ RANS simulations. With average runtimes of approximately 1.0 h at $\alpha=2^\circ$ and 1.5 h at $\alpha=10^\circ$ on a 32-core node, this corresponds to 2605 node-hours, or 83,360 core-hours. By contrast, the active multi-fidelity campaign required $154\times1.0+99\times1.5=302.5$ node-hours, or 9,680 core-hours, for the RANS portion of the workflow. These values are intended to quantify the reduction in RANS usage relative to the same fixed automated high-fidelity pipeline; they should not be interpreted as a universal cost ratio between XFOIL and all possible RANS implementations.

While CST offers a compact and flexible parameterization, the physical role of individual coefficients is not immediately interpretable. To relate design variables to aerodynamic behavior, a set of geometric descriptors was extracted for each airfoil: maximum thickness and its location ($t_{max}, x_{t_{max}}$), maximum camber and its location ($C_{max}, x_{C_{max}}$), leading-edge radius ($LE_r$), trailing-edge angle ($TE_\alpha$), and peak/average curvature for both camber ($\ddot{C}_{max}, \overline{\ddot C}$) and thickness ($\ddot{t}_{max}, \overline{\ddot t}$). A Pearson correlation analysis between the CST parameters ($\theta$), geometric metrics, and aerodynamic cost values ($J, E, C_L$) is shown in \autoref{fig:09_corr}.
The dominant correlations indicate that performance variations ($E, C_L$, and the aggregate cost $J$) are primarily associated with camber-related features. Both the maximum camber magnitude ($C_{max}$) and its chordwise position ($x_{C_{max}}$) exhibit a strong positive correlation with the aerodynamic performance. This aligns with fundamental aerodynamic theory, identifying camber as the primary mechanism for lift maximization, a trend clearly reflected in the high-$C_L$ peak observed in the optimization history and the Pareto front structure. Conversely, thickness-related measures (e.g., enclosed area $A$) correlate negatively with $E$, consistent with the tendency of the optimizer to reduce thickness in order to mitigate drag penalties at cruise. These trends provide an interpretable link between the Pareto-front geometry families discussed previously and the scalar objective driving selection.

A useful composite descriptor is the asymmetry ratio $\zeta = C_{max}/t_{max}$, which compactly summarizes the relative dominance of camber over thickness. The observed correlation of $\zeta$ with the aerodynamic objectives indicates that high-performing designs in this campaign tend to increase camber while keeping thickness moderate, consistent with the efficiency-driven penalty of excessive thickness at cruise. At the parameter level, $\theta_{9-11}$ exhibits the strongest linear associations with the dominant geometric descriptors, indicating that these coefficients govern much of the performance-relevant variability within the chosen CST bounds. In contrast, $\theta_{3}$ shows a more selective influence, correlating primarily with camber-related features. The remaining CST coefficients display weaker correlations with the cost function, suggesting a lower first-order sensitivity for the specific operating conditions and objective definition considered here.

\begin{figure}[t]
    \centering
    \includegraphics[width=0.95\linewidth]{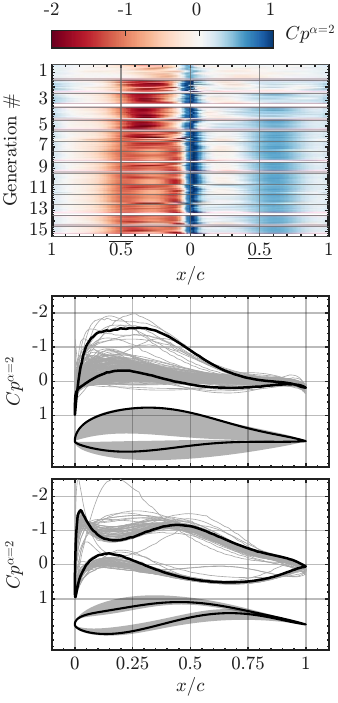}
    \caption{Cruise pressure-coefficient evolution, showing the full-population unwrapped $C_p$ map, first and converged (8) generation shape and $C_p$ statistics.}
    \label{fig:10_cp_2}
\end{figure}

\begin{figure}[t]
    \centering
    \includegraphics[width=0.95\linewidth]{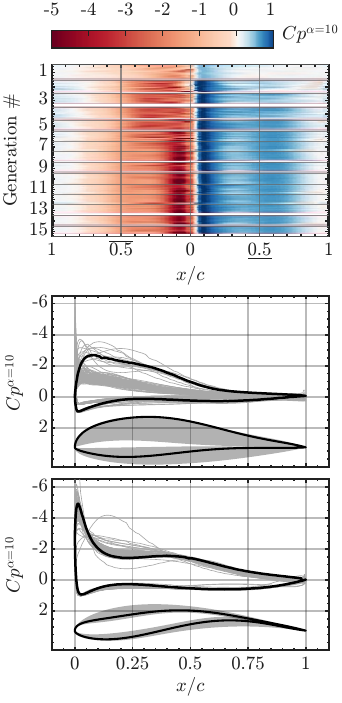}
    \caption{Take-off pressure-coefficient evolution, showing the full-population unwrapped $C_p$ map, first and converged (8) generation shape and $C_p$ statistics.}
    \label{fig:10_cp_10}
\end{figure}

To provide physical insight into the optimization trends, the pressure coefficient ($C_p$) distributions for all evaluated individuals are summarized in \autoref{fig:10_cp_2} and \autoref{fig:10_cp_10} top panels using an unwrapped representation, where each row corresponds to one individual and the chordwise coordinate is mapped along the horizontal axis. This visualization highlights that the two operating points promote distinct loading strategies. At cruise ($\alpha=2^\circ$, \autoref{fig:10_cp_2}), improvements in $E$ are associated with a redistribution of suction and a moderation of pressure recovery, which is consistent with reducing drag penalties linked to adverse pressure gradients and separation sensitivity.
At take-off ($\alpha=10^\circ$, \autoref{fig:10_cp_10}), the optimization favors higher integrated loading, achieved by strengthening the suction-side depression and increasing pressure-side loading while maintaining a pressure recovery compatible with attached flow. Together, these trends illustrate the coupled compromise enforced by the composite objective across the two conditions.

The bottom panels of \autoref{fig:10_cp_2} and \autoref{fig:10_cp_10} compare representative $C_p$ profiles at Generation 1 and at Generation 8 (after the best design has stabilized under the HyGO stopping criterion). For cruise ($\alpha=2^\circ$, \autoref{fig:10_cp_2}), the final design exhibits a more front-to-mid-chord distributed loading than the early-generation best candidate. In particular, the suction-side profile shows a pronounced leading-edge depression followed by an extended region of relatively weak recovery up to approximately mid-chord, yielding a sustained pressure difference ($C_{p,\ell}-C_{p,u}$) over the fore–mid chord. This smoother recovery is consistent with limiting mid-chord adverse-pressure-gradient severity in a highly cambered configuration, which is favorable for maintaining attachment and controlling drag. 
By contrast, the best Generation 1 profile is more characteristic of thickness-driven improvement, with a milder suction peak and a more conservative chordwise evolution of $C_p$, consistent with the early tendency of the optimizer to exploit thickness increases to raise aerodynamic performance.

At take-off ($\alpha=10^\circ$, \autoref{fig:10_cp_10}), early improvements similarly increase the pressure difference through a combination of stronger suction-side depression and increased pressure-side loading, consistent with lift augmentation. The final best design, however, shifts the emphasis toward camber redistribution at moderate thickness: the selected $C_p$ profile combines a strong leading-edge suction peak with an extended region of comparatively gentle recovery on the suction side, reducing recovery steepness around mid-chord ($x/c\approx0.5$) where separation onset is typically sensitive.
Overall, the population-level maps and the representative profiles support the transition inferred from the Pareto and correlation analyses: coarse thickness variation dominates early exploration, whereas later improvements are achieved primarily through camber shaping and suction redistribution subject to separation-aware pressure recovery constraints.

\begin{figure}
    \centering
    \includegraphics[width=8.cm]{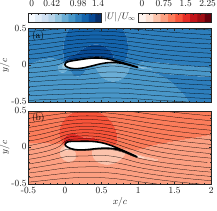}
    \caption{RANS velocity contours and streamlines for the optimized airfoil at cruise and take-off, showing predominantly attached flow at both operating conditions.}
    \label{fig:11_flow_best_gen1}
\end{figure}
\autoref{fig:11_flow_best_gen1} complements the $C_p$-based interpretation by providing a qualitative view of the near-body flow for the final optimized airfoil at the two operating points. At $\alpha=2^\circ$, the velocity field remains smooth around the leading edge and through the aft region, with streamlines following the suction-side curvature without visible separation, supporting the conclusion that the efficiency gains are achieved through load redistribution and controlled recovery rather than through operating near a separated state. At $\alpha=10^\circ$, the field shows the expected acceleration over the suction side associated with the high-$C_L$ solution while preserving attached streamlines, indicating that the final geometry sustains the required lift through increased suction and favorable recovery without inducing large-scale separation. Given the 2D RANS context, these visualizations should be interpreted as qualitative evidence of attached/fully separated behavior (rather than a definitive characterization of transitional or small laminar separation bubbles), but they are consistent with the solver-admissibility checks and with the “separation-aware” recovery trends previously inferred from the population $C_p$ evolution.

\section{Conclusions}\label{s:Conclusions}
This work presented an active multi-fidelity framework for airfoil shape optimization that combines inexpensive low-fidelity evaluations from XFOIL with sparse, computationally intensive high-fidelity RANS simulations. A low-fidelity-informed Gaussian process regressor was used to map geometric variables and low-fidelity aerodynamic outputs to high-fidelity performance metrics, thereby mitigating systematic low-fidelity biases and providing predictive uncertainty estimates that can be exploited during the search.

An uncertainty-triggered refinement strategy was introduced to adaptively select candidates for high-fidelity evaluation, concentrating computational effort in regions where the surrogate is insufficiently reliable. By enforcing a prescribed uncertainty threshold on surrogate-based predictions and iteratively assimilating newly acquired high-fidelity data, the surrogate accuracy is progressively improved in the portions of the design space explored by the optimizer, particularly in the vicinity of the identified minima. The condition-wise decoupling of surrogate models further enables multi-point optimization without requiring high-fidelity evaluation across the full operating set when uncertainty is localized to a subset of conditions.

The framework was coupled with HyGO, a hybrid genetic algorithm, to retain the global search capabilities of evolutionary optimization while promoting model refinement. To limit the accumulation and propagation of surrogate-induced errors, a modified elitism strategy was proposed in which elite candidates are systematically validated using the high-fidelity solver. The resulting high-fidelity data are used to update the surrogate and to synchronize the population costs before reproduction, improving the reliability of the information presented to the optimizer. A diversity constraint on elite selection further prevents redundant high-fidelity evaluations and mitigates premature concentration of the search around a single basin of attraction.

The proposed methodology was demonstrated on a two-point optimization representative of cruise ($\alpha=2^\circ$) and take-off ($\alpha=10^\circ$) at $Re=6\times 10^{6}$. Under these conditions, the optimized design achieved simultaneous improvements of $20.75\%$ in take-off lift coefficient ($C_L^{\alpha=10^\circ}$) and $41.05\%$ in cruise aerodynamic efficiency ($E^{\alpha=2^\circ}$). In addition, the pre-update validation results confirm that the LF-informed transfer surrogate substantially reduces the LF–HF discrepancy for the drag-sensitive efficiency objective while requiring only limited correction for $C_L$ at take-off, which is consistent with the observed asymmetry in uncertainty-triggered high-fidelity calls across operating points.

Overall, the main contribution of this study is an optimization-embedded active multi-fidelity strategy in which (i) high-fidelity escalation is governed by predictive uncertainty at the candidate level rather than by a fixed per-generation budget, and (ii) synchronization mechanisms (elite validation, population re-evaluation, and diversity-aware elitism) are used to control surrogate-induced drift within an evolutionary optimizer. The conclusions drawn in this work remain conditional on the assumptions inherent to the chosen solvers and problem formulation, the use of XFOIL/RANS as the low-/high-fidelity pair, the CST parametrization, and the two-condition two-dimensional airfoil optimization problem considered. For larger mission profiles involving many operating conditions, the condition-wise decoupled formulation may require sparse or multi-output Gaussian-process models, shared latent representations, or clustering of operating conditions to preserve scalability. Similarly, extension to three-dimensional wings or multidisciplinary configurations would require adapting the feature representation, meshing/solver workflow, and surrogate architecture. While the proposed pipeline is transferable to other configurations and is expected to reduce computational cost, the resulting savings would depend on the specific automated RANS workflow adopted as the high-fidelity reference, including mesh resolution, convergence criteria, and solver settings. Consequently, alternative high-fidelity discretizations would change the absolute runtime comparison but not the candidate-level fidelity-escalation and synchronization mechanisms introduced in this work. Nevertheless, the results support the use of online-refined multi-fidelity surrogates coupled with global, model-free optimizers for multi-point aerodynamic shape optimization. Future work will focus on extending the framework to richer fidelity hierarchies, broader operating envelopes, additional constraint sets, and three-dimensional or multidisciplinary configurations.


\section*{Funding Sources}
This work has been supported by the TIFON project, ref. PLEC2023-010251/ MCIN/AEI/ 10.13039/501100011033, funded by the Spanish State Research Agency. Oriol Lehmkuhl has been partially supported by a Ramón y Cajal postdoctoral contract (Ref: RYC2018-025949-I). We also acknowledge the Barcelona Supercomputing Center for awarding us access to the MareNostrum V machine based in Barcelona, Spain.

\section*{Code and Data availability}
The framework will be made openly available to the research and engineering community after the review process is finished, under the MIT license, including source code, documentation, and ready-to-use examples. All libraries employed for the algorithm development are open-source, and the baseline optimization software employed can be found in \href{https://github.com/ipatazas/HYGO}{HyGO}.

\bibliographystyle{elsarticle-num-names} 
\bibliography{biblio.bib}

\appendix
\section{Static-surrogate optimization}\label{anex:static_GPR}

To isolate the effect of the online surrogate-update mechanism, an additional ablation study was performed using a static LF-informed GPR. The same initial high-fidelity training set, LF-informed surrogate formulation, scalar objective, and HyGO hyperparameters were retained, but the surrogate was kept fixed after initialization. Therefore, no uncertainty-triggered high-fidelity enrichment, mandatory high-fidelity elite validation, or population synchronization was applied during the optimization. This setup represents a common offline surrogate-assisted workflow in which a surrogate is trained from an initial design of experiments and then optimized without further high-fidelity correction.

\begin{figure}
\centering
\includegraphics[width=8.cm]{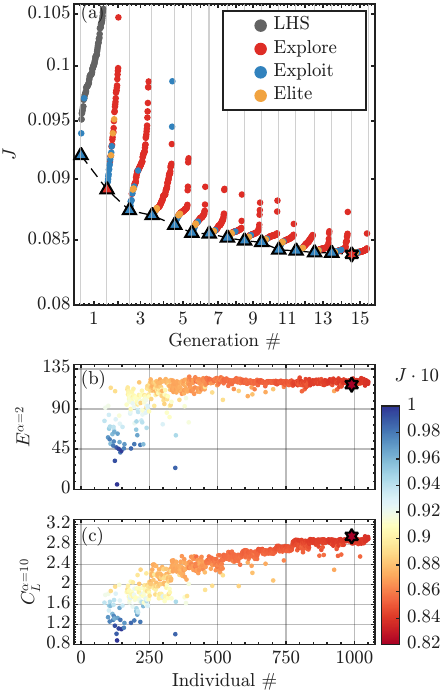}
\caption[]{Static-GPR ablation results showing surrogate-predicted cost convergence, generation operators, cruise efficiency, take-off lift, and overall best design.}
\label{fig:05_no_rans_J_evo}
\end{figure}

\begin{figure}
    \centering
    \includegraphics[width=7.5cm]{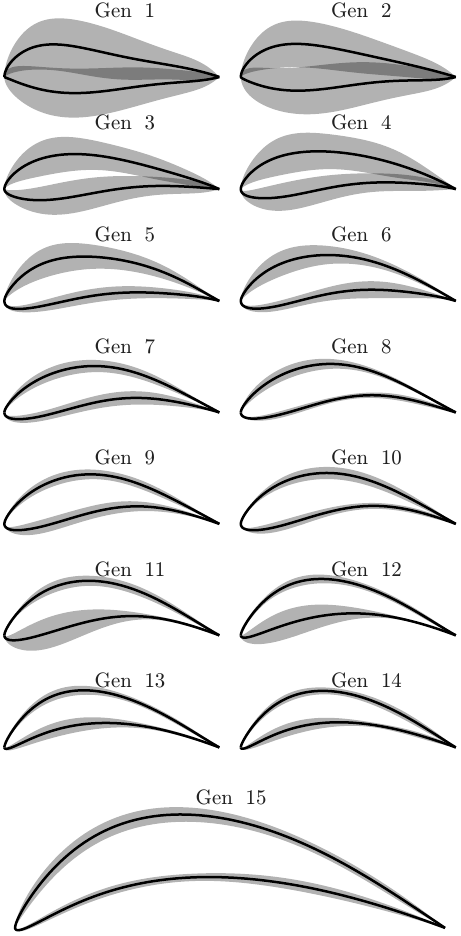}
    \caption[Geometric evolution of the static-GPR ablation]{Static-GPR ablation geometry evolution, showing generation-wise mean airfoils over the corresponding sampled-geometry envelopes.}
    \label{fig:052_J_evolution_airfoils_no_rans}
\end{figure}

The obtained results are displayed in \autoref{fig:05_no_rans_J_evo}. The static surrogate produces an apparently lower predicted cost than the online-refined strategy and exhibits a smoother optimization profile, mainly due to the monotonic increase in the predicted take-off lift shown in \autoref{fig:05_no_rans_J_evo}(c). However, post hoc RANS validation reveals that this apparent improvement corresponds to a surrogate-induced false optimum. The selected geometry exhibits exaggerated camber, as shown by the final generations in \autoref{fig:052_J_evolution_airfoils_no_rans}, and lies outside the effective support of the initial RANS training set. For this candidate, the static surrogate predicts $E^{\alpha=2^\circ}=117.24$, whereas the RANS value is only $E^{\alpha=2^\circ}=20.92$, corresponding to an overprediction of approximately $460\%$. Similarly, the predicted take-off lift is $C_L^{\alpha=10^\circ}=2.96$, whereas the RANS value is $C_L^{\alpha=10^\circ}=2.235$, corresponding to an error of approximately $32.6\%$. This discrepancy is mainly associated with separated-flow behavior induced by the excessive camber, which the static surrogate fails to capture.

\begin{figure}
    \centering
    \includegraphics[width=8.cm]{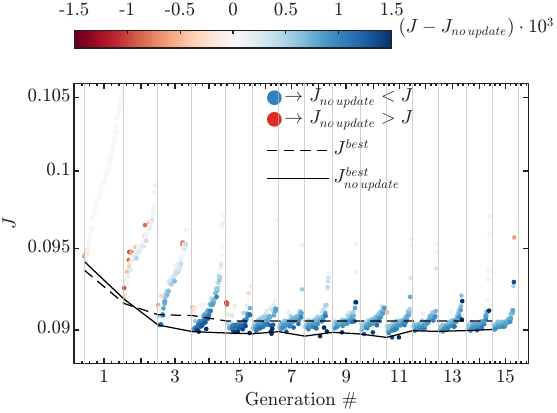}
    \caption[Comparison between online-refined and static-GPR optimization paths]{Comparison of online-refined and static-GPR optimization paths, showing cost differences, static-run ordering, and best-cost histories for both strategies.}
    \label{fig:10_comparison_GPR}
\end{figure}

The comparison between the geometric evolution of the online-refined optimization in \autoref{fig:05_J_evolution_airfoils} and the costs that would be obtained with no online GPR update is illustrated in \autoref{fig:10_comparison_GPR}. Without high-fidelity enrichment and synchronization, the optimizer progressively exploits surrogate predictions in a region of increasing camber and thickness, where the initial LF-informed GPR is poorly supported by RANS data. In contrast, the proposed online-refined strategy corrects the surrogate as the search evolves, preventing the optimizer from propagating geometries whose predicted performance is not reliable under high-fidelity validation, avoiding unrealistic separation-prone airfoils. 

The ablation, therefore, illustrates the role of the proposed online refinement and synchronization mechanisms. The objective of the dynamic high-fidelity sampling strategy is not only to reduce the number of RANS evaluations but also to prevent the optimizer from exploiting surrogate predictions in regions that are insufficiently supported by high-fidelity data. In the present problem, the static surrogate drives the search toward a non-validated low-cost region, whereas the proposed method uses uncertainty-triggered enrichment, high-fidelity elite validation, and synchronization to maintain the optimization trajectory within regions where the surrogate remains consistent with RANS evaluations.

This ablation should be interpreted as a diagnostic comparison rather than as a comprehensive benchmark of active-learning strategies. Fixed-budget enrichment, expected-improvement infill, adaptive Kriging, and related sampling policies may provide alternative ways to allocate high-fidelity evaluations, and a systematic comparison among such policies is outside the scope of the present study. The purpose of the static-GPR baseline is narrower: it isolates the effect of removing online high-fidelity correction, elite validation, and synchronized re-evaluation from the same optimizer and initial surrogate. Under this controlled removal, the optimizer converges toward a RANS-inconsistent false optimum, which is precisely the failure mode addressed by the proposed framework.

\end{document}

%% file: utils/packages.tex
\usepackage{textcomp}

\usepackage[version=4]{mhchem}
\usepackage{siunitx}
\usepackage{graphicx} 
\usepackage{subcaption}
\usepackage{tcolorbox}
\usepackage{tabularray}
\usepackage{array}
\usepackage{makecell}
\usepackage{longtable,tabularx}
\usepackage{colortbl}
\usepackage{hhline}   
\usepackage{multirow} 
\usepackage{booktabs}
\usepackage{amsmath}  
\usepackage{amsfonts}   
\usepackage{mathtools}
\usepackage{bm}
\usepackage{subfiles}
\usepackage{blindtext}
\usepackage{minitoc}
\usepackage{setspace}
\usepackage{endnotes} 
\usepackage{fancyvrb} 
\usepackage{rotating}

\usepackage{dcolumn}
\usepackage[utf8]{inputenc}
\usepackage[T1]{fontenc}
\usepackage{etoolbox}

\usepackage{algorithmic}
\usepackage{float}

\usepackage{xcolor}
\usepackage{tikz}

\usepackage{epstopdf, epsfig}
\usepackage{calc}
\usepackage{microtype}
\usepackage{lscape}
\usepackage{esint,upgreek}
\usepackage{afterpage}
\usepackage{listings}

\usepackage{framed} 
\usepackage{multicol} 
\usepackage{nomencl} 

\usepackage[linesnumbered,ruled,vlined]{algorithm2e}
\RestyleAlgo{ruled}

\DontPrintSemicolon
\SetKwComment{Comment}{$\triangleright$\ }{}
\SetKwInOut{KwRequire}{Require}
\SetKwInOut{KwEnsure}{Ensure}
\SetCommentSty{itshape}
\SetKwComment{tcp}{}{}

\usepackage[normalem]{ulem}

\usepackage[numbers]{natbib}


%% file: utils/defined_colors_and_commands.tex
\usepackage[normalem]{ulem}
\usepackage{color}
\usepackage{utils/mylines}
\usepackage{utils/myplots}
\newcommand{\sy}[2]{\mbox{(\kern-.25em\SymbolRGB[solid]{#1}{.8pt}{#2}{5pt}\kern-.25em)}}
\newcommand{\lsy}[3]{\mbox{(\kern-.1em\lineSymbolRGB{#1}{#2}{2pt}{#3}{4pt}\kern-.45em)}}
\newcommand{\lcap}[2]{~\,{\kern-1em\protect\mylcap{#1}{#2}}}

\definecolor{blue}{rgb}{0,0,1}
\definecolor{red}{rgb}{1,0,0}
\definecolor{black}{rgb}{0,0,0}
\definecolor{white}{rgb}{1,1,1}
\definecolor{greyR}{RGB}{50,50,50}
\definecolor{redR}{RGB}{227, 47.3333, 39}
\definecolor{greenR}{RGB}{55, 160.3333, 85}
\definecolor{blueR}{RGB}{55, 135, 192.3333}
\definecolor{yellowR}{rgb}{0.9412, 0.7843, 0.0588}

\definecolor{greyRr}{RGB}{188,188,188}
\definecolor{redRr}{RGB}{239, 182, 182}
\definecolor{blueRr}{RGB}{179, 223, 241}
\definecolor{greenRr}{RGB}{180, 245, 217}

\definecolor{yellowelite}{rgb}{0.9961, 0.6774, 0.2167}
\definecolor{blueelite}{rgb}{0.192156862745098,	0.509803921568627,	0.741176470588235}
\definecolor{greenregion}{rgb}{0.0925443054757589	0.451731373410683	0.203847910313637}
\definecolor{purpleregion}{rgb}{0.574518222139030	0.392066460638973	0.643779710468552}

\definecolor{myorange}{HTML}{D95319}
\definecolor{myblue}{HTML}{0072BD}
\definecolor{myyellow}{HTML}{EDB120}
\definecolor{mygreen}{HTML}{77AC30}

\definecolor{color_rebut}{RGB}{20, 20, 255}
